\documentclass{aastex}
\usepackage{emulateapj5}
\usepackage{epsfig}

\def\arcsec{$^{\prime\prime}$}
\def\arcmin{$^{\prime}$}
\def\micron{$\mu$m}
\def\msolar{M$_{\odot}$}

\def\lsim{\mathrel{\lower .85ex\hbox{\rlap{$\sim$}\raise
.95ex\hbox{$<$} }}}
\def\gsim{\mathrel{\lower .80ex\hbox{\rlap{$\sim$}\raise
.90ex\hbox{$>$} }}}

\begin{document}
{\title{EVIDENCE OF HALO MICROLENSING IN M31$^{1,2}$}
\footnotetext[1]{Data were obtained in part using the 1.3 m McGraw-Hill Telescope of the MDM Observatory}
\footnotetext[2]{Based in part on observations with the VATT: the Alice P.~Lennon Telescope and the Thomas J.~Bannan Astrophysics Facility.}}
\author{Robert R.~Uglesich$^{3,4}$, Arlin P.S.~Crotts$^{3,5}$,
Edward A. Baltz$^{6,7}$, Jelte de Jong$^8$, Richard P.~Boyle$^9$,
Christopher J.~Corbally$^9$}
\affil{
3.\ Department of Astronomy, Columbia University, 550 W.~120th St., New York,
NY 10027, U.S.A.;\newline
4.\ Department of Biomathematics, Mt Sinai School of Medicine, NY, NY 10029,
U.S.A.;\newline
5.\ Institute for Advanced Study, Olden Lane, Princeton, NJ, 08540,
U.S.A.;\newline
6.\  ISCAP, Columbia Astrophysics Laboratory, 550 W.~120th St., New York, NY
10027, U.S.A.;\newline
7.\ KIPAC, 2575 Sand Hill Road, Menlo Park, CA 94025, U.S.A.;
8.\ Kapteyn Astronomical Institute, Postbus 800, Groningen ~9700 AV,
Netherlands;
9.\ Vatican Observatory Research Group, Steward Observatory, University of
Arizona, Tucson, AZ ~85721, U.S.A.  }

\begin{abstract}
We have completed an intensive monitoring program of two fields on either
side of the center of M31, and report here on the results concerning
microlensing of stars in M31.
These results stem from a three-year study (the VATT/Columbia survey) of
microlensing and variability of M31 stars, emphasizing microlensing events of
3 day to 2 month timescales and likely due to masses in M31.
These observations were conducted intensively from 1997-1999, with baselines
1995-present, at the Vatican Advanced Technology Telescope and the 1.3-meter
telescope at MDM Observatory, with additional data from the Isaac Newton
Telescope, including about 200 epochs total.
The two fields monitored cover 560 square arcminutes total, positioned along
the minor axis on either side of M31.
Candidate microlensing events are subject to a number of tests discussed here
with the purpose of distinguishing microlensing from variable star activity.
A total of four probable microlensing events, when compared to carefully
computed event rate and efficiency models, indicate a marginally significant
microlensing activity above that expected for the stars alone in M31 (and the
Galaxy) acting as lenses.
A maximum likelihood analysis of the distribution of events in timescale and
across the face of M31 indicate a microlensing dark matter halo fraction
consistent with that seen in our Galaxy towards the Large Magellanic Cloud
(Alcock et al.~2000a).
Specifically, for a nearly singular isothermal sphere model, we find
a microlensing halo mass fraction $f_b=0.29^{+0.30}_{-0.13}$ of the total dark
matter, and a poorly constrained lensing component mass (0.02 to $1.5 M_\odot$,
1 $\sigma$ limits).
This study serves as the prototype for a larger study approaching completion;
between the two there is significant evidence for an asymmetry in the
distribution of microlensing events across the face of M31, and therefore a
large population of halo microlensing dark matter objects.
\end{abstract}


\section{Introduction}

\bigskip

The nature of the dark matter in the halo of disk galaxies remains a mystery.
While this component of galaxies contributes the majority of their mass, most
other data about the dark matter are indirect at best.
Some clue as to characteristics of part of the dark matter may be implied by
the microlensing detection of masses towards the Large Magellanic Cloud.
The MACHO survey revealed a frequency of microlensing that was unanticipated
in the context of the visible stellar population alone, but still falls short
of accounting for the entire dark matter halo (Alcock et al.~2000a).
In particular, the survey indicates a most probable dark matter halo mass
fraction of 20\% (limits of 5\% to 50\%, at the 95\% confidence level), for an
indicative spatial distribution model (singular, isotropic, isothermal sphere).
Constraints from the EROS magellanic cloud survey (Afonso et al.~2003) yields
a constraint consistent with 20\% microlensing halo fraction, but also
consistent with no microlensing halo.
There are few independent indications that bear on the validity of the
presence of a microlensing halo component.
Fluctuations in the brightnesses of images of lensed QSO MG~0414+0534 indicate
a significant but sub-dominant halo fraction of microlensing masses (Schechter
\& Wambsganss 2002).

A decade ago we proposed that M31 offers a favorable alternative
venue for probing the halo dark matter problem in spiral galaxies, by applying
the microlensing techniques to stars in M31 itself, for lenses primarily in
M31 but also the Galaxy.
In particular, such a signal could be easily distinguished in terms of an
asymmetry in microlensing rate across the face of M31, and could be monitored
effectively using image subtraction to suppress the severe crowding of M31's
stars (Crotts 1992).
This required developing techniques for the application of image subtraction to
a time series of images (Tomaney \& Crotts 1996), which led to the first
candidate microlensing events in M31 (Crotts \& Tomaney 1996).
At least several more years of such observations have been required to both
amass sufficient lensing events for a statistically meaningful sample and to
cull out variable stars by extending the baseline.
The current work presents the results from this more extended survey and offers
our interpretation of the microlensing observations which have resulted.
While our survey has found several thousand variable stars, to be reported
elsewhere, we have also isolated a sample of events that are more consistent
with microlensing events, and imply that a significant fraction of the dark
matter halo in M31 is due to objects of stellar mass.

M31 microlensing surveys have also been conducted by the AGAPE groups (Baillon
et al.~1993, Calchi Novati et al.~2002), and the survey reported here has been
extended by MEGA (Crotts et al.~2001, deJong et al.~2003),
the first results of which we mention in conjunction with these in reaching our
conclusion.

\section{Observations}

\begin{figure}
\centerline{\psfig{file=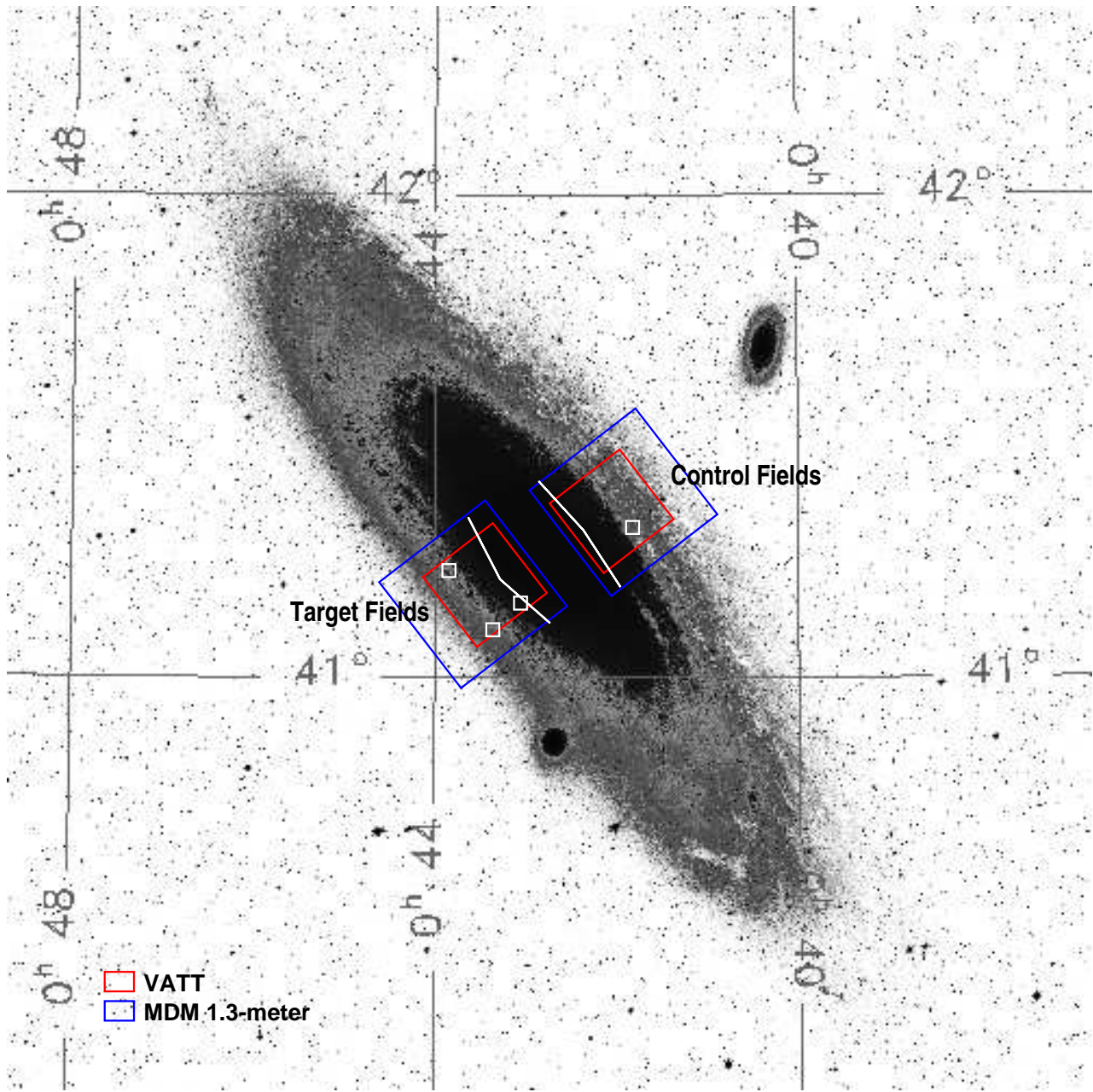,height=17cm}}
\caption[M31 (NGC 224): The Andromeda Galaxy]{
Ground-based optical image of M31 taken from the Digitized Sky Survey (DSS).
The locations of the nearside (``Control'') and farside (``Target'') fields are
indicated for the VATT (smaller FOV) and MDM 1.3-meter telescopes.
The positions of the four surviving microlensing candidates are indicated by
the smallest squares (left to right: 99-3688, 97-1267, 97-3230 and 99C-1259,
respectively).
The INT/WFC fields are shown elsewhere (de Jong et al.~2003).
The blackened portion of the galaxy shows where M31 surface brightness
typically dominates over foreground sky.
The white curves show the approximate boundary where events are ignored due
to lower photometric efficiency (see \S 5).
Image size is approximately $2^{\circ}.2$ square.
}
\label{fig:m31}
\end{figure}

Our data consist of a long time sequence of images taken of two fields: farside
and nearside (see Table 1 and Figure \ref{fig:m31}).
The positions of our fields were chosen to maximize the number of stars being
monitored and to provide optimal leverage for measuring the near to far-side
optical depth asymmetry (\S 5): two fields along the minor
axis which are on opposite sides of the nucleus and at equal galactocentric
radius, largely just beyond the bulge.
The southeastern field (``Target'' or farside) farther from Earth than the
nucleus of M31, on the disk in the region of highest expected optical depth
(Crotts 1992).
The northwestern (``Control'') field lies on the nearside of the disk, closer
to Earth than the nucleus.
The fields are rotated by $37^{\circ}.7$ to align them with M31's principal
axes.
The inner edge placement is chosen to minimize the detector area which is
saturated by the nuclear light.
This offset from the nucleus is approximately 0\arcmin.5 and 1\arcmin.5 and the
fields span the radii 0.11-3.91 kpc (0.49-17.38 kpc, in projection) and
0.33-2.84 kpc (1.47-12.62 kpc, in projection) for MDM and VATT, respectively
(assuming a distance to M31 of 770 kpc).
The R-band surface brightness in our fields ranges from 16.9 mag arcsec$^{-2}$
to 22.2 mag arcsec$^{-2}$ (Walterbos \& Kennicutt 1988).
Comparison with sky brightness as a function of lunar phase (Walker \& Smith
1999) indicates that, except for the outermost portions of our fields, M31
surface brightness dominates over the sky for all lunar phases, which allows us
to make use of telescope bright time which is much easier to acquire.

The observations made for this survey were obtained at three telescopes:

\subsection{Vatican Advanced Technology Telescope (VATT)}
The Vatican Advanced Technology Telescope (VATT) on Mt.~Graham, Arizona is an
aplanatic, f/9 Gregorian system featuring a fast, $f/1.0$, 1.8-meter primary
mirror, and is designed to frequently obtain subarcsecond seeing.
Over our three years of our observing on the VATT for this project, we measure
the median seeing at VATT to be 1\arcsec.09.  
The imaging camera (known as the Columbia SSCCD) featured a SITe
$2048\times2048$ (24 \micron) pixel$^{2}$ thinned, backside-illuminated CCD
covering a square field of view (FOV) of 11\arcmin.3 on a side, with CCD gain
and readnoise of 2.5 e$^{-}$/ADU and 9 e$^{-}$, respectively, and linear
response over $0-28000$ ADU.  
We installed a doublet biconvex achromat corrector lens that produces uniform
20 \micron~(0.\arcsec25) diameter spots and best focus over the entire surface
corresponding to the curved (approximately 220 \micron~sagitta center-to-edge)
backside-illuminated face of the CCD.
This flat focal plane allows for relatively straightforward image subtraction
to be performed (\S 3.1).

\subsection{ MDM 1.3-meter (McGraw-Hill) Telescope}
The MDM 1.3-meter (McGraw-Hill) telescope on the southwest ridge of Kitt Peak,
Arizona is an $f/7.6$, 1.32-meter Ritchey-Creti\'en system.
The camera utilized was an MDM facility imager known as Echelle, featuring a
SITe $2048\times2048$ pixel$^{2}$ CCD nearly identical to the SSCCD and
covering a field of view of 17\arcmin.0$\times$17\arcmin.0.
The median seeing of our MDM data, spanning the time period August 1997 -
December 1999, is 1\arcsec.65.
Significant astigmatism and comatic aberration are present in the optical
system and the telescope focus can be variable (25\% changes in FWHM on
timescales of 10-15 minutes on occasion).
Frequent monitoring of the focus allowed us to maintain image quality.

\subsection{Isaac Newton Telescope}
Additional images were obtained at the Isaac Newton Telescope (INT
2.5-meter), La Palma, Canary Islands, with the Wide Field Camera (WFC).
The INT/WFC data are detailed elsewhere (de Jong et al.~2003).
For this telescope the fields are configured in a north-south pair rather than
on either side of the projected major axis of M31, but almost all of the VATT
fields are covered.
\bigskip

The CCD pixel scale for the VATT, MDM 1.3-meter and INT correspond to
0.330, 0.497 and 0.332 arcsec, respectively.
Even in the best seeing, these images are oversampled with respect to Nyquist
frequency.

For the VATT and MDM images, 
survey filters were chosen to maximize sensitivity to red giants which
constitute the dominant stellar population in our fields ($<V-I> = +1.2$:
Tonry 1991) and also to provide color
separation in order to test for achromaticity of microlensing candidates.
We have employed non-standard R and I filters (designated R$_{jt}$ and
I$_{custom}$) which are slightly broader than their Cousins equivalents
(Cousins 1974).
R$_{jt}$ has a more uniform response than R$_{cousins}$ across the bandpass
which extends from $\lambda$5700 (just beyond the [O {\sc i}] $\lambda$5777
night sky line) to $\lambda$7100 (just short of the atmospheric {\sc OH-}
emission feature), and I$_{custom}$ extends from $\lambda$7300 to
$\lambda$10300 (5\% power points).
For INT/WFC images, the standard $r^\prime$ and $i^\prime$ filters, close to
those in the SDSS system (Fukugita et al.~1996) were chosen, since R$_{jt}$ and
I$_{custom}$ were unavailable there.
This filter system is similar to the R$_{jt}$ and I$_{custom}$, except that the
filters in the SDSS system cross over in transmission approximately 200\AA\
redder than in our system, and the $i^\prime$ filter extends only to about
8700\AA\ in the IR, where CCD detector response is falling rapidly.

Since our primary sieves for microlensing events involved the R-band data,
some priority was given to R over I-band.
Furthermore, data was sometimes completed in the farside field at the expense
of coverage in the nearside field.
For this work, all images taken in the same field and filter on the same night
were combined into a nightly stack.
These are summarized in Table 2.
(To save table space, in some cases we merge consecutive nights onto the same
line when different field/filter combinations were used.)~
The stacking procedure, and further analyses, are detailed below.

We emphasize here that the efficiency calculation to compare predicted and
measured event rates, and the microlensing candidate Criteria I-V
listed below, were preformed using the MDM data alone, with VATT and INT data
being included in the analysis in the last stages of candidate selection.

\section{Image Analysis}

\subsection{Difference Image Photometry (DIP) Pipeline}
\begin{figure}
\centerline{\psfig{file=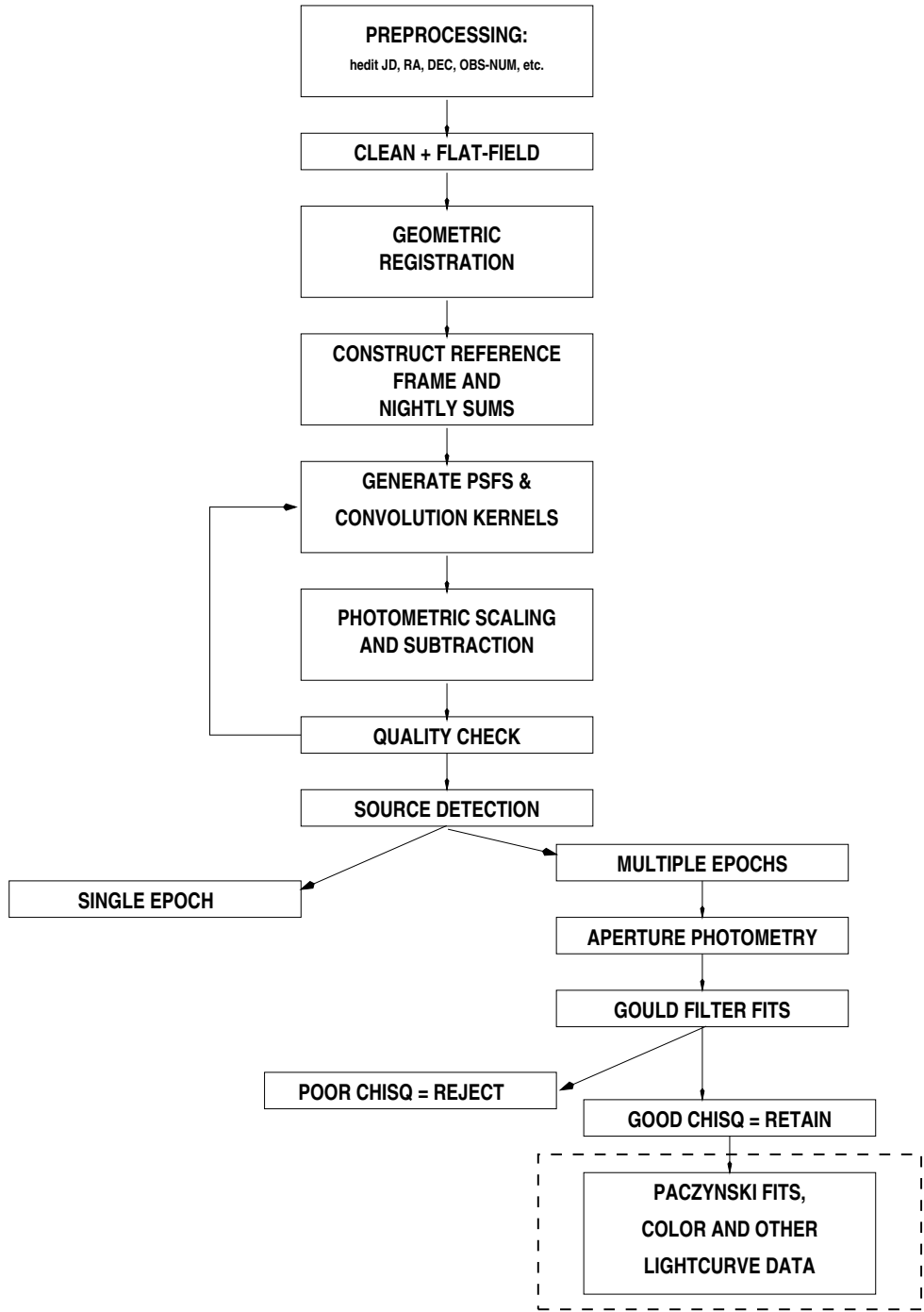,height=25cm}}
\vskip -1.2in
\caption
{A schematic representation of the sequence of steps involved in processing the
survey data with the Difference Image Photometry pipeline. 
The step outlined by the dashed box includes information from other sources as
detailed in the text.
\bigskip
\bigskip
}
\label{fig:flow}
\end{figure}

The data reduction pipeline, excluding the source filtering and lightcurve
fitting routines, has been constructed to function as a package, named
DIFIMPHOT, using the Image Reduction and Analysis Facility
(IRAF).\footnote{IRAF is distributed by the National Optical Astronomy
Observatories, which are operated by the Association of Universities for
Research in Astronomy, Inc., under cooperative agreement with the National
Science Foundation.} 
The sequence of steps involved in the pipeline are outlined in Figure
\ref{fig:flow}. 
Briefly, the procedure involves geometric registration of images to a common
coordinate system, construction and application of a convolution kernel to
correct for the spatial and temporal variability of the point-spread-function
(PSF), accurate photometric scaling followed by subtraction, and aperture
photometry of point sources detected in the subtracted image. 
The procedure is explained in detail in this section. 

\subsection{Preliminary Processing}
As with any large observational survey, adopting a uniform standard for
managing data is essential. 
The survey data were obtained at various observatories each of which,
unfortunately, employed a different FITS header format and which did not always
contain all of the information of interest to us. 
This was particularly a problem at VATT as the SSCCD camera could not
communicate with the telescope control computer. 
We began by ensuring that every image header contained the following
information: right ascension and declination in J2000 coordinates, epoch,
filter, exposure time, UT date and time, Julian date, field label (target or
control), and a unique 11-digit identifying integer. 
Airmass and parallactic angle information were also added.

\subsection{ Flat-fielding and Cleaning}\label{FF+Clean}
All frames were debiased and flat-fielded in a standard manner. 
The bias level was determined from the column overscan region except when the
bias frames exhibited residual 2-D structure after overscan subtraction. 
On those nights, a median average of the full, 2048$\times$2048 pixel, bias
frames was used.  

Sky-flats in each bandpass, smoothed with a median filter of width
$\sim$~75$\times$ PSF FWHM, were used to divide out the overall illumination of
the CCD. 
Dome-flats were used to divide out the pixel-to-pixel variation. 
Sky-flats at VATT were taken without the field corrector in order to minimize
internal reflections. 
Consequently, flat-fielding errors of a few percent were present in VATT images
on spatial scales much larger than the PSF. 
This contributed to the background noise level but did not affect photometric
scaling which was calculated using the ``unsharp masked'' frames (see
Section~\ref{sec-stack}). 
We could not circumvent this problem by using super-sky-flats constructed from
the M31 exposures themselves, given the uneven illumination of the CCDs and
the repetitive way in which these observations must be obtained.

Bad pixels and cosmic rays were identified on each frame by fitting a
5$\times$5 pixel 2-D (elliptical) Gaussian of width slightly less than the
seeing to each pixel in the image and computing the difference between the data
and the model. 
A pixel associated with a CCD defect or cosmic ray hit is easily discernible
because it is poorly fit by a Gaussian and will produce large residuals over
the fit region. 
Replacement values were calculated by simple linear interpolation across the
narrowest dimension spanning the bad pixels. 
Removing cosmic rays early was crucial because, once the PSF matching
convolution kernel was applied (Section~\ref{sec-psfmatch}), these defects
would have become similar to stellar PSFs and be potentially cataloged as
sources in the difference frame.

Frequently, we checked the noise and gain properties of the CCDs using a
dataset consisting of a pair of bias frames and a pair of dome-flats,
unprocessed so that the noise properties were not altered, input to FINDGAIN to
estimate the gain and read-noise of the CCD using the algorithm of Janesick
(1987). 
If multiple sets existed, then the values of the gain and read-noise were taken
to be the median of the individual estimates. 
This information was saved in the image headers.

\subsection{ Registration and Stacking}
\subsubsection{ Geometric Registration of Images}
In order to maximize the overlap area covered by the individual exposures, the
image representing the median pointing for a particular field (target or
control) during the first season of observations was chosen as the reference
for that particular telescope+field combination. 
All frames were registered to their corresponding common coordinate system
({\it i.e.,} target or control).  
This was done by first measuring the centroids, using the PHOT routine, of many
(N$_{*}\sim 100$ for VATT, and $\sim$250 for MDM data) bright, unsaturated and
isolated (no companions or saturated regions within $r \leq$~5$\times$ PSF
FWHM) stars on the median pointing frame to generate a reference coordinate
list. 
There were a number of marginally resolved M31 globular clusters present in our
survey fields and care was taken to exclude them from this list. 
Next, the central 512$\times$512 pixel region of each unregistered frame was
searched for bright stars and these (typically $\sim$25 stars) were matched to
a subset of the reference list sources using an automated algorithm developed
(Groth 1986) which searches for similar triangles formed by triplets
of points in each list. 
From this, a crude coordinate transformation, consisting only of translation
and rotation terms, was obtained and used to transform the reference list to
the coordinate system of the unregistered frame. 
These coordinates were accurate enough that, when input to the PHOT routine,
the sources were recovered on the unregistered frame and precise centroids were
measured. 
This paired list of reference and unregistered coordinates was input to the
IRAF routine GEOMAP which fit a 4$^{th}$-order polynomial with half cross terms
in order to determine the full geometric transformation.  

GEOMAP computed linear and distortion term separately. 
The linear term includes an $x$ and $y$ shift, an $x$ and $y$ scaling, a
rotation and a skew. 
The distortion term consists of a polynomial fit to the residuals of the linear term. 
Legendre polynomials were used for the fits to the higher-order terms as they
tend to provide more stable solutions than power series polynomials.  
The images were then transformed using the GEOTRAN routine with bicubic spline
interpolation and flux conserved, in the standard manner, by multiplying the
registered pixel values by the Jacobian of the coordinate transformation. 
The final registration was accurate to better than 0.15 pixels RMS, and all
frames had PSFs with FWHM $\geq$2.15 pixels which allowed for adequate
sampling, and minimized resampling errors and corresponding degradation of the
PSF. 

\subsubsection{ Construction of Image Stacks}\label{sec-stack} 
The next step in the reduction process was to create stacks, by combining
individual exposures on various timescales, to achieve increased 
signal-to-noise ratio $S/N$. 
We were interested primarily in creating nightly stacks, as well as, a single,
high $S/N$ (adding insignificantly to the total error), good seeing reference
stack for each observing season. 
While the restrictions on data quality (e.g. maximum allowed seeing, scaling,
airmass etc.)\ differed for these two types of stacks, the general procedure
followed was the same for both.

The first step in combining frames was to ``unsharp mask'' the data. 
This removed the underlying smoothed M31 and sky background and was done by
constructing a large-scale median filtered frame which was subtracted from the
raw frame.
Median filtering is a computationally expensive calculation and, instead of
applying a filter of large diameter, we chose to initially apply an 8$\times$8
pixel boxcar average to the data and then median filter that frame with a
filter of 7$\times$7 to achieve results similar to a median on larger spatial
scales. 
For a 2048$\times$2048 pixel$^{2}$ image, the boxcar+median is nearly
100 times faster to calculate (~$\overline{t}\sim12$ sec versus $1000$ sec)
and produces a result nearly indistinguishable from a straight application of a
median filter.

On every frame under consideration for inclusion in the stack, fluxes were
measured at the positions of the stars belonging to the reference coordinate
list described above. 
One frame, usually representing the best seeing and lowest sky background, was
designated as the photometric reference and all photometric scalings were
calculated relative to the fluxes measured on this frame. 
A separate estimate of the scaling was calculated for each star and the overall
scaling for a given frame was taken to be the median of these values, after
rejecting those stars which were found to vary in flux by more than 5\%. 
Images with poor seeing, high scale values and high background levels could be
excluded from consideration.  The images were assigned weights, $w$, which were
proportional to the inverse variance of the noise in a seeing element
\begin{equation}
w = \frac{C}{FWHM^{2} \times scale^{2}} \;\; ; \;\; \sum_{n} w_{i} = 1
\end{equation}

Combined frames have altered noise characteristics that must be tracked in
order to assign meaningful errors to our photometric measurements (Section
\ref{sec-diphot}). 
For each image stack, consisting of $n$ exposures with gain $g$ [e$^{-}$/ADU],
read-noise $r$ [e$^{-}$], fractional flat-field error $p$, signal $s$ [ADU],
and weight $w$, the effective quantities are calculated in the following manner
and recorded in the image header.

\begin{eqnarray}
G & = & \frac{\sum_{n} s_{i}g_{i}w_{i}}{\sum_{n} w_{i}} \nonumber \\
R & = & \sqrt{\frac{\sum_{n} (s_{i}^{2}/g_{i}^{2}) r_{i}^{2} w_{i}}{\sum_{n} w_{i}}} \nonumber \\
P & = & \sqrt{\frac{\sum_{n} s_{i}^{2} p_{i}^{2} w_{i}}{\sum_{n} w_{i}}}
\end{eqnarray} 

The properties of the individual stacks are given in Table 2.

\subsection{ Image Subtraction: PSF Matching the Fourier
Way}\label{sec-psfmatch}
Crotts (1992) suggested that it might be possible to search for
variability in unresolved stellar fields by subtracting two images separated in
time. 
On the timescales of interest (minutes to many months) the majority of stars
are photometrically stable and the subtracted frame will have a smooth, zero
mean background with isolated positive and negative point sources at the
location of stars which are intrinsically variable. 
In principle, the idea is elegant and straight-forward to implement but, in
practice, it is hampered by the fact that the point spread function of the
telescope is spatially and temporally variable due to changes in the
atmospheric and site conditions (seeing) and focus of the telescope. 
These PSFs changes must be measured and corrected. 
Here we describe a technique based on Fourier convolution (as opposed to the
alternative method of image subtraction based on image decomposition into
linearized basis functions: Alard \& Lupton [1998] which gives comparable
performance),
first utilized by Ciardullo et al.~(1990) to search for novae in M31
globular clusters by comparing frames taken in broad and narrow bands.
Our implementation (Tomaney \& Crotts 1996) concentrates on variability, and
successfully compensates for PSF variations and produces subtracted images with
a background which is less than twice the theoretical photon noise limit. 

Consider a given region on two images each having a different point spread
function, $r$ and $i$, but which can be related to each other by the following
convolution with kernel $k$ 
\begin{equation}
i = r * k.
\end{equation}
By the convolution theorem,
\begin{equation}
I = R \times K
\end{equation}
where $I$, $R$ and $K$ are simply the Fourier transforms of $i$, $r$ and $k$,
respectively, and, it follows, that $k$ can simply be represented as
\begin{equation}
k = {\cal F}^{-1}\left[\frac{R}{I}\right]
\end{equation}
where ${\cal F}^{-1}$ is the inverse Fourier transform.

This is an idealized description and our data do not have infinite $S/N$, so we
are careful to not introduce spurious artifacts into the convolution kernel $k$
during the Fourier transform procedure. 
In particular, the low $S/N$ wings of the PSF which are located at large radii
map to the central core (small spatial frequency, $\lambda=1/r$) of the PSF in
Fourier space and, to minimize this contamination, we fit these wings below
some threshold (defined as a percentage of the peak flux) with a 2D elliptical
Gaussian.  

We divide each of our frames into $n$ $\times$ $n$ subregions over which we are
certain that the PSF is not significantly spatially varying. 
Typically, $n=4$ for MDM data and can be as low as $n=1$ (a single PSF
accurately describing the entire field of view) for VATT data, taken with the
special corrector lens. 
In each subregion, we identify as many resolved and isolated stars as possible
and extract the $51\times51$ pixel region centered on each star. 
These data are resampled onto a common coordinate system, flux normalized and
median combined after sigma-clipping to remove any spurious pixels. 
By constructing an empirical PSF in this fashion, we improve the $S/N$ over
that obtained using only a single star per subregion and this allows us to drop
our elliptical Gaussian model replacement threshold to levels below 3\% of the
peak flux. 
This is especially crucial for the MDM data where PSFs at large field radii
exhibit comatic aberration which is not accurately modeled by an elliptical
Gaussian.
The data are written to a multi-extension FITS file containing $n^{2}$ image
planes corresponding to the $n^{2}$ PSFs.
Next, the Fourier transforms for each corresponding pair of empirical PSFs are
computed via an FFT algorithm. 
We are interested in the convolution which will degrade the better seeing frame
to the poorer seeing one and the quotient is constructed maintaining this
sense. 
It is conceivable that this sense is not identical for all subregions of an
image. 
The quotients are inverse Fourier transformed and then written to a FITS file
with $n^{2}$ image plane.

The matching and photometric scaling prior to subtraction is done in a
smoothed, piecewise fashion. 
For each subregion, the appropriate image is convolved with the kernel
generating a PSF matched subregion. 
A photometric scaling between the two matched images is calculated using the
fluxes of the same resolved and isolated stars used to construct the empirical
PSF. 
After scaling, the two images are subtracted from each other.
The scale factor information is written to the image header as it modifies the
effective gain which is used to calculate the expected Poisson noise in the
difference frame.

As described in Section \ref{sec-stack}, we construct a high $S/N$, good-seeing
stack which spans an entire observing season as well as nightly image stacks. 
The high $S/N$ reference frame is {\em always} subtracted from each of the
nightly stacks.

\subsection{ Source Detection and Photometry}\label{sec-diphot}

\begin{figure}
\centerline{\psfig{file=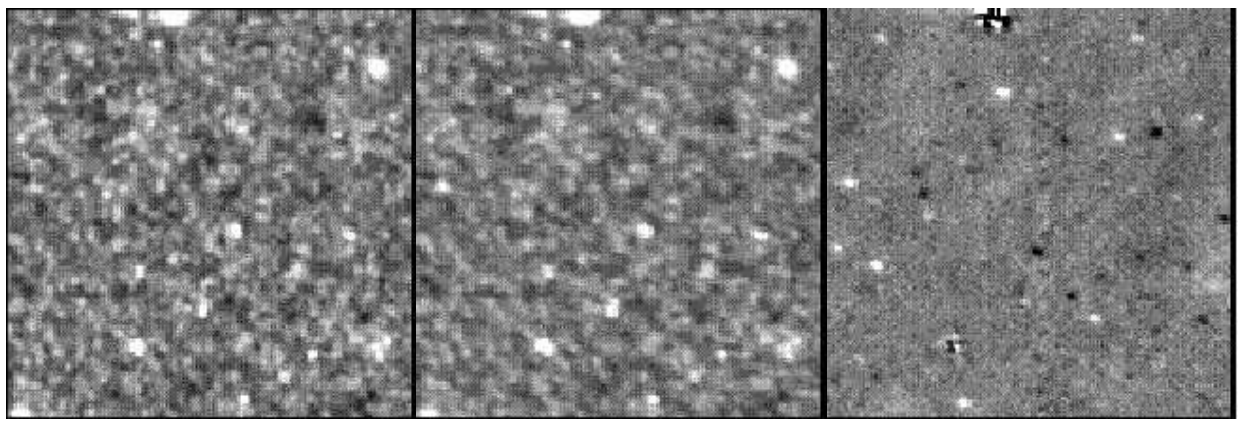,height=30cm}}
\vskip -4.0in
\caption
{This figure demonstrates a typical result obtained using our image subtraction
procedure. 
The panel to the left is a 1\arcmin\, $\times$ 1\arcmin\, subregion taken
from near the center of the reference frame for the 1998 observing season. 
The center panel is the same subregion from a nightly stack taken 393 days
earlier. 
The rightmost panel shows the result of image subtraction. 
Notice that most of the sources seen in the difference frame are not apparent
to the eye in the upper panels.  All data are from the MDM 1.3-meter telescope.
}
\label{fig:imsub}
\end{figure}

The final data products that are generated by the pipeline are lightcurves of
the detected variable sources. 
As one can see from Figure \ref{fig:imsub}, in the center of the fields the
residuals of variables are well isolated and sufficiently far from the
crowded-field regime that straightforward aperture photometry can be performed. 
(However, at the inner edge of the field, where surface brightnesses are
several magnitudes higher, this is not so.
See \S 5.)~ 
The first step in this process is to create a catalog containing the locations
of these positive and negative residuals and we would like to accomplish this
in some automated fashion searching for objects whose peak fluxes are more than
a specified number of standard deviations above or below the background.

Point sources are detected using the SExtractor software package (Bertin \&
Arnouts 1996) SExtractor searches for local density maxima which are composed
of some minimum number of connected points, all of which are at least some
specified number of standard deviations above the background. 
The power of this approach is that the software detects objects without making
{\it a priori} assumptions about their shapes. 
This is useful for finding objects whose PSFs are not strictly Gaussian such as
the stars which suffer from coma in our MDM images. 
SExtractor computes the local standard deviation of the background which is
crucial in dealing with our data because the dominant background noise source
is the surface brightness of the disk which exhibits a gradient across the
field and by using local noise estimates we are able to set a single $S/N$
threshold for the entire frame (as opposed to the procedure using DAOPHOT
(Stetson 1987).

Before searching for sources, we mask out residuals associated with diffraction
spikes, border discontinuities created by the registration process and
residuals resulting from poor subtraction of bright, resolved stars (e.g.,
stars like the one located at the top edge of Figure \ref{fig:imsub}). 
The bright stars are easily masked out by taking the list of stars used in the
construction of the geometric registration solution or in the construction of
the empirical PSFs and zeroing the pixel values in a circular region of radius
$5\times$ FWHM centered on each star. 
The diffraction spikes and border effects are not as prevalent and are removed
by hand. 

All frames were searched for both positive and negative point sources which
were at least 4-$\sigma$ above the background and which were composed of at
least ten contiguous pixels. 
The source lists were then culled by requiring all sources to have at
least two detections anywhere in the time series. 
We have cataloged more than 8000 unique sources in our MDM survey data.

The next task is to perform aperture photometry on these point sources. 
This is done using an optimal extraction algorithm (Naylor 1998) which rejects
pixels whose values differ from a fit to the empirical PSF by more than a
specified number of $\sigma$ and replaces them with the model. 
Flux is measured within an aperture of radius = 1.5 $\times$ FWHM. 
The aperture correction applied is calculated as the fraction of total flux to
the flux within the same size aperture as measured for the empirical PSF
corresponding to that particular subregion on the frame.
The sky background is taken to be the mode of the values in an annulus of width
0.5 $\times$ FWHM with inner radius at 2 $\times$ FWHM.

Additional points from the INT/WFC sample were added to lightcurves after
events were identified from the R-band lightcurves from the VATT and MDM
samples.

\section{Distinguishing Microlensing Events from Variable Stars in M31}

The analyses above are sufficient to isolate a variable source from the huge
density of stellar sources in ground-based images of M31.
We still need techniques to discriminate microlensing events from the usual,
more numerous (or more rare) ways in which stars vary intrinsically.

Earlier microlensing surveys based on resolved sources (MACHO, EROS, OGLE,
MOA, which nonetheless suffer from source blending) have established the
reality of the microlensing phenomenon and
tested the validity of several characteristics which are easily applied to
distinguish microlensing events from other types of variations seen in stellar
populations.
This characteristics include, for individual events of point sources and point
masses
\footnote {
The critical surface density required for a mass to be well described as a
point lens corresponds to $\sim10^4- 10^5$~g~cm$^{-2}$ for source-lens
distances as migth be found here.
For any of the stars considered here this density is exceeded.
}

\bigskip
\noindent
1) the microlensing amplification should be wavelength-neutral, in the sense
that the lensing event produces the same factor enhancement at the same time
at all wavelengths across the spectrum, given the fact that photons interact
identically with the lens gravitational field regardless of photon energy,

\bigskip
\noindent
2) the lensing events should follow a paczynski curve (Paczynski 1986) i.e.,
observed flux
$f(t) = f_{0} \, A[u(t)]$,
where the amplification is
$A = (u^2+2)/[u(u^2+4)^{1/2}]$,
and the impact parameter normalized to the einstein radius is
$u(t) = r/R_e = [u_0^2 + v^2 (t-t_0)^2]^{1/2}$.

\bigskip
\noindent
3) the event should effectively never repeat, since the event duration is usual
much longer than the product of the survey duration and the microlensing
optical depth.

Once enough events have been amassed, further tests can also be applied to the
whole sample:

\bigskip
\noindent
4) a population of microlensed sources should be randomly drawn from the
available sources e.g., they should not favor any particular region in the
color-magnitude diagram, and 

\bigskip
\noindent
5) the events should also exhibit randomly distributed impact parameters $u_0$,
modulated only by observational selection biases on $u_0$, since the relative
projected positions of source and lens should be realized at random.

\bigskip
In M31, we have difficulty establishing (1) directly, since being forced to
remove the baseline flux from the lightcurve we eliminate knowledge of the
actual amplification.
Nonetheless, the residual flux $(1-A)f_0$ should also be wavelength-neutral,
providing a less rigorous but still useful test.
Condition (5) is also applied with great difficulty, since fits of paczynski
curves to $(1-A)f_0$ are nearly degenerate in $A_{max}=A(u_0)$ and hence $u_0$,
which can be traded off against unknown parameters $v_0  = v/R_e$ or lens
mass $m$, unless very high signal-to-noise ratio $S/N$ data of the event is
collected, particular in the wings of the peak (Baltz \& Silk 2000).

In M31, however, we have the added benefit of easily extending our search over
large portions of the galaxy, and therefore have the potential of
distinguishing microlensing events from false backgrounds on the basis of
spatial distribution.
This effect produces two related criteria for distinguishing halo microlensing
events from false backgrounds due to variable stars:

\bigskip
\noindent
6.a) the strong farside/nearside asymmetry (Crotts 1992), due to the higher
density of lenses and more favorable lensing geometry for stars on the more
distant side of M31's disk, or

\bigskip
\noindent
6.b) the expectation of a diminished gradient in microlensing events versus
stellar populations as one moves away from the center of the galaxy (Gyuk \&
Crotts 2000, Baltz, Gyuk \& Crotts 2003), since the distance from the lens to
the source $D_{ls}$ grows roughly linearly with the projected distance from the
center of M31.
Because of this the optical depth grows, as $D_{ls}$, like the microlensing
cross-section $\sigma_{lens} = \pi R_e^2 = 4 \pi D_{ls} D_{ol} /G m D_{os}$,
where $D_{ol} \approx D_{os}$ in the case of M31.

\subsection{Preliminary Filters}

We have developed a set of automated procedures which, when applied in
succession, will repeatably extract a set of microlensing candidates from our
full lightcurve database. 
The first steps towards isolating microlensing from stellar variability deal
with the better-sampled $R$ lightcurves.
The first few step, in order of application, include:

\bigskip
Criterion
I. Minimum number of detections: for a lightcurve to be catalogued, detections
of at least $4 \sigma$ in 2 nightly stacks must occur at the same point (to
within 1.0 arcsec).
This produces a sample of 8162 lightcurves.
\bigskip

In principle this sample serves as the repository of microlensing events which
are not well-approximated by paczynski curves, such as binary-mass or
planetary lenses, which we treat in a separate analysis.
For now we will make the simplifying but somewhat undercautious assumption that
the preponderance of events are due to simple point masses (and sources), as
found in surveys towards the Galactic Bulge and Magellanic Clouds.

\bigskip
Criterion II.
Good fit to Gould filter function:
In a differential analysis, the lightcurves represent the quantity
$\Delta f = f(t) - f_{0} = f_{0} \, [A(t) - 1]$
which can also be fit for the same four parameters $f_0$, $u_0$, $v_0$ and
$t_0$.
For unresolved source stars we cannot recover $f_{0}$ accurately from the data
and the fits are nearly degenerate in the parameters $u_0$ and $f_{0}$.
Gould (1996) has proposed an alternative functional description which,
using fewer parameters, adequately describes a microlensing event in this
unresolved regime.
In the limit $u_0 << 1$, the excess flux takes the form 
$f_{0} \, [A(t) - 1] \rightarrow \frac{f_{0}}{u_0} \, G(t)$,
where
$G(t) ~ \equiv ~ [\omega_{\rm eff}^{2} (t - t_{max})^{2} + 1]^{-1/2}
\omega_{\rm eff} ~ \equiv ~ \frac{\omega}{u_0} \nonumber 
\omega ~ \equiv ~ t_{0}^{-1}$.
This ``Gould filter function'' $G$ allows us to describe a lensing event of an
unresolved star with only three parameters, $f_{0}/{u_0}$, $t_0$, and
$\omega_{eff}$. 
We can estimate the extent to which a lensing event lightcurve is adequately
described by $G$ by constructing the correlation, $\eta(b)$, between $G$ and
$(A-1)$:
$$\eta(u_0) = \frac{\int_{-\infty}^{\infty} dt ~ [A(t) - 1] ~
G(t)}{[\int_{-\infty}^{\infty} dt ~ [A(t) - 1]^{2}]^{1/2}
~ [\int_{-\infty}^{\infty} dt ~ G(t)^{2}]^{1/2}}.$$
This fit deviates most significantly from the true microlensing amplification
profile in the wings of the event leading us to adopt a slightly loose
threshold for the goodness-of-fit, $\chi^{2}/\nu < 20$. 
Microlensing events are well represented by this Gould filter function $G$ and
we retain only those lightcurves below the $\chi^{2}$ threshold. 
Additionally, we require that the fit maximum be contained in the range of
lightcurve points in $R_{JT}$ found within a FWHM of the peak.
These requirements retain exactly 100 lightcurves.

The vast majority of our variable sources are periodic in time and are fit
poorly by $G$, however; later we take further steps to make sure that a
constant flux baseline is maintained by a source beyond the event well-fit as
microlensing.

\bigskip
Criterion III.
Adequate sampling of event maxima: 
Since we seek rare events, we limit our detection microlensing candidate peaks
to the well-sampled portion of our time series of observation by requiring good
sampling during the peak of a candidate microlensing event. 
We retain only candidates which have at least 4 points during the event with
flux difference greater than 4-$\sigma$ above the baseline and which lie on
both sides of the peak, based on 1.3-meter $R$-band data only.
After this requirement 45 lightcurves remain.
\bigskip

Until this point we have instituted a series of criteria which are likely to
reject not only variable stars, but also poorly sampled or inadequately
detected true microlensing events.
These first three criteria greatly influence the detection efficiency of our
survey, therefore, and are factors modelled in our theoretical calculation
of these efficiencies, discussed in \S 5.

Next, we fit the $R$ lightcurves of these well-sampled events with a paczynski
curve, with five fit parameters ($u_0$, $v_0$, $t_0$, baseline flux, flux
offset).
The offset flux parameter arises from the possibility that the reference image
includes flux from epochs when the star was not at baseline, hence resulting in
flux subtracted from the star's signal in individual epochs.
For this and all succeeding criteria, the expectation is that an insignificant
fraction of true microlensing events should be rejected.

\bigskip
Criterion IV.
MDM 1.3-meter, $R$-band paczynski curve fit must be less than $\chi^2/\nu = 2$.
This threshold will rule out less than one in
$10^4$ of true lensing events for lightcurves with the number of points as
ours, much less than one event for the sample surviving criteria I - III, upon
which our efficiency calculation is based.
This criterion actually reduces the surviving candidates to 26.
\bigskip

From this point, and with a much smaller sample of lightcurves, we move on to
establish more rigorous filters against variable stars entering our candidate
microlensing sample.
Some of these tests are reminiscent of others applied in Galactic microlensing
surveys, but attention must be paid due to differences in M31: the ignorance of
the baseline flux due to crowding in ground-based images (dealt with via image
subtraction), the brighter (and redder) population of stars being used as
sources, and the intrinsic differences in populations between this portion of
M31 and the Magellanic Clouds (and to a lesser extent the Bulge).
Furthermore, we bring in further data, from the $I$-band and from VATT and
INT, which serve to eliminate variable stars, but are used in a way to leave
the number of true microlensing events substantially unchanged.

\subsection{Microlensing \& Variability: M31 versus Galactic Searches}

At first, studies of microlensing in the Magellanic Clouds seemed to benefit
from the advantage of studying a population relatively free of any known,
threatening populations of variable objects.
None were perviously recognized to flare occasionally in a way easily confused
with a microlensing event e.g., nearly wavelength-neutral and similar in shape
to a paczynski curve.
Nonetheless, after one year's survey, a previously unknown population fitting
this description was discovered (Cook et al.~1994, Keller et al.~2002),
known as ``aperiodic blue variables'' or ``bumpers''.

In M31 we can use the large-scale spatial distribution of microlensing events
versus variable stars to distinguish mock lensing events from real ones, but
suffer from the condition of a known, common population of variable stars
which mimic microlensing events to an irritating degree.
This is due to our requirement to concentrate on brighter, therefore often
redder, stars many of which show troublesome levels of variability.
Here we describe methods whereby such variables can be discriminated from
microlensing events.
Since the primary culprits are miras and other red supergiant variable stars,
of which there are thousands in our fields, we must establish a filter to
sieve these from our sample with great certainty.

\begin{figure}
\epsfig{file=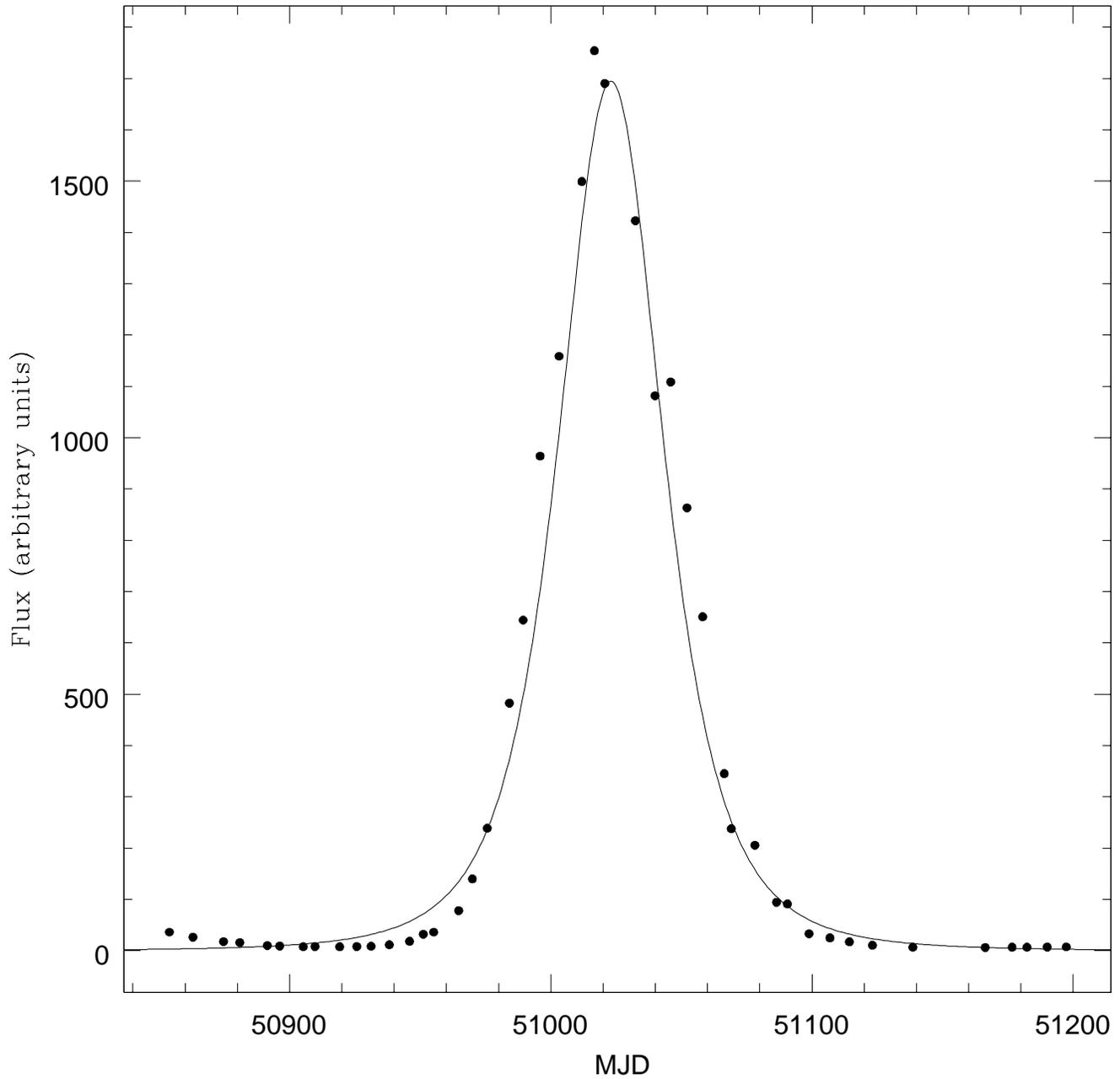,width=\columnwidth}
\caption{
Lightcurve (in roughly V magnitude) of T UMa as observed by VSOLJ
(points), and fit (solid trace) by a paczynski curve (with $u_0 = 3$).
The deviations between the two curves amount to only 6\% of the maximum light
amplitude.
T Uma has a period of $256.6^d$, so the data plotted cover 1.6 periods.
}
\label{fig:tuma}
\end{figure}

We have warned for some time (Uglesich et al.~1997) that miras are capable of
mimicking paczynski curves given insufficient sampling and $S/N$ in the data.
This is illustrated in Figure \ref{fig:tuma}, which shows the lightcurve for
the mira-type variables T UMa (period$ = 256.6^d$, amplitude of variation $A
\approx 5^{mag}$) which we selected from the VSOLJ (2003) sample purely on the
basis of large quantity of data and appropriate period.
Not all miras undergo such a good fit, especially since a large fraction are
significantly asymmetric in maxima.
Data are taken in roughly the V-band, and fit by a paczynski curve with impact
parameter $u_0 = r_0/r_{ein} = 3$ and velocity $v_0 = 0.09d^{-1}$.
The average residuals from this fit (due primarily to imperfect fit rather than
measurement error) correspond to only 6\% of the maximum light flux.
The FWHM of this mira lightpulse is 67d, which corresponds to the einstein
timescale of a halo object mass in our M31 field of roughly $1 ~M_\odot$.
With a period of roughly 2/3 year, a mira such as this observed one year later
(or previously) will remain within 5\% of the baseline for 126d, or nearly the
entire length of an M31 observing season (typically $\sim$150d,
August to January).
Thus a paczynski-like variation, observed in one band, must be observed either
at extremely high $S/N$ or over a minimum of three M31 seasons in order to be
distinguished from a mira.

The comparison of lightcurve shape in two bands is likewise insufficient to
distinguish strongly against miras, since the colors e.g., $R-I$ as used here,
can vary little across the maximum lightpulse.
For instance, we take a sample of 12 mira variables (de Laverny et al.~1997)
observed in $UBV(RI)_c$ and compute the variation in color over the maximum
pulse to find that in many cases the difference in $R$ and $I$ lightcurves,
once normalized to one another, varies by less than 10\% of the peak $R$ flux,
over the peak (within 2.5 magnitudes of maximum).
In particular, once we bin points of similar phase in order to reduce scatter
due to measurement error (and perhaps fluctuations between cycles), we find
that for 4 of the 12 stars (R Oct, RU Oct, V Cha and X Hyi), the adoption of a
best-fit $R-I$ color allows one to predict at all times the flux in one band
given the flux in the other, consistently to within 5\% of the peak flux
(or standard deviation of 3.7\%).
From this we conclude that there exists a large population of mira variables
for which individual object's lightcurve shapes in $R$ and $I$ are
indistinguishable from each another without very high $S/N$ data (total event
detection on the order of 50 $\sigma$), hence they cannot be distinguished from
microlensing events on the basis of the wavelength-neutrality criterion.

With the failure of two of the primary means for distinguishing microlensing
events from a large class of variable stars, we strive to construct additional
filters to remove miras and other variables from our sample.
We consider additional known populations of variables, but are cautious, given
the size and coverage of our dataset compared to previous studies, to the
possibility that undiscovered contaminants might slip into our microlensing
event sample due to unknown populations in this portion of M31.
Sufficiently rare contaminants might be easily ruled out as having a different
spatial distribution from that expected for microlensing events.

\begin{figure}
\epsfig{file=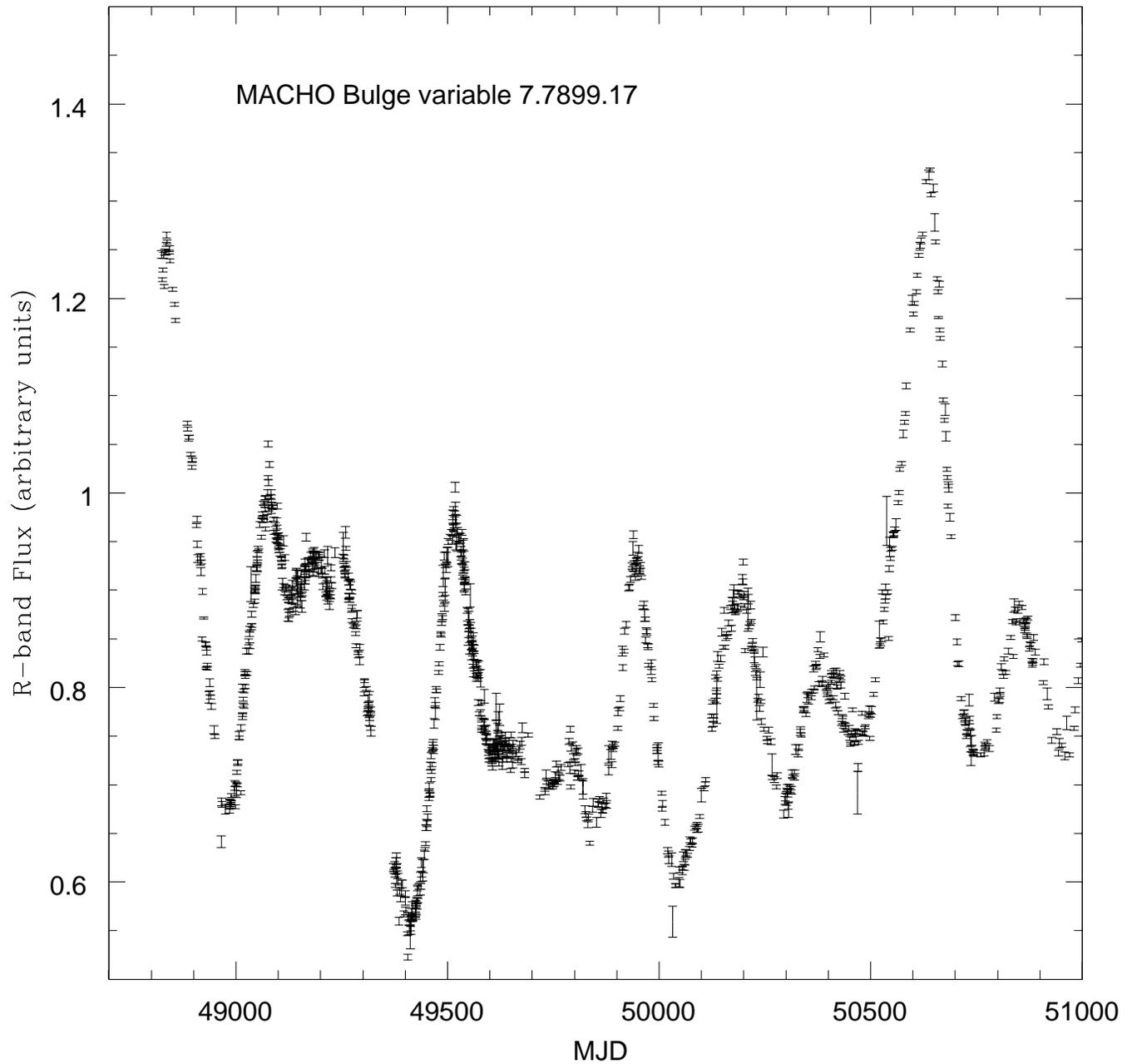,width=\columnwidth}
\caption{
A superlative example of a semi-regular type variable with an isolated, 
high-amplitude peak which might be mistaken for a microlensing event.
}
\label{fig:machovar}
\end{figure}

While Criteria I and II largely fail to exclude miras, Criterion IV should
succeed if coverage is sufficient to detect any plausible periodicity.
As described above, this demands at least three well-sampled seasons of data,
as achieved here.
Semiregular variables (SRs) are more numerous than miras, by a factor of 20 in
the Galactic Bulge (Alard et al.~2001).
Miras are separated from SRs at an amplitude of variation $A = 2.5$ mag (for
lightcurves in $R$), with SRs usually much less variable than this threshold.
(For SRs in Alard et al., the median $A$ is 0.2 mag, even though very few $A$
values are less than 0.1 in that sample.
The largest $A$ value for SRs in that sample is 1.0 mag.)~
The usual behavior of SRs is to vary nearly continuously over amplitudes which
are large but not overwhelming compared to the baseline flux.
In terms of confusion with microlensing events, SRs contrasts favorably with
miras, which can hide most of their cycle within a few percent of maximum flux
near the baseline then suddenly undergo a positive pulse.
These general SR properties do not imply, however, that positive deviations in
SR lightcurves are always small.
A perusal (Alves 2002) of the MACHO variable star database (Welch 2002)
reveals the most extreme case of a variable (Figure \ref{fig:machovar}) which
can remain constant in flux to within $\pm$20\% for several years, then spike
to a $+$70\% maximum, in this case extending over 107d FWHM.
\footnote{
This paper utilizes public domain data obtained by the MACHO Project, jointly
funded by the US Department of Energy through the University of California,
Lawren ce Livermore National Laboratory under contract
No. W-7405-Eng-48, by the National Science Foundation through the Center for
Particle Astrophysics of the University of California under cooperative
agreement AST-8809616, and by the Mount Stromlo and Siding
Spring Observatory, part of the Australian National University.}
While such instances are rare in the Bulge, their frequency in the inner
disk/outer bulge of M31 is unknown.
Infrequently sampled data might confuse them with a microlensing event.
Rather than try to discriminate against such isolated spike behavior on the
basis of amassing sufficient time coverage, we excluded such stars as follows
by eliminating the entire class which might be subject to such variations.

We deal with these mira and SR variables by employing a variant of
characterisitic
(5) above, requiring that source stars rest at a locus in the color-magnitude
diagram that is not occupied by a population of sufficiently variable stars.
Aperiodic blue variables typically have $-3.5 \la M_V \la -1.5$ (Keller et
al.~2002) and blue colors $V - I \la 0.3$.
Red supergiant variables are more difficult to isolate; we use variability of
such stars in the Galactic Bulge and LMC to prescribe color-magnitude regions
to avoid.

The MACHO Project variable star catalog was searched both for LMC and Bulge
variables which have large enough variations to pass our threshold in $R$.
The catalog has a entry each for baseline magnitude, amplitude of variation,
and variation timescale.
We take the latter two to indicate a characteristic deviation from the baseline
and flag a star as variable if it passes our R detection threshold (which
corresponds to unit amplitude on a 1-day timescale for a source of $R=19.3$).
Since the catalog also contains $V-I$ colors, we can plot these variable
sources on a color-magnitude diagram (Figure \ref{fig:cmd}).
This database did not include cepheids.
To trace the instability strip we include cepheids from the LMC (Caldwell
\& Coulson 1985) and the Galaxy (Feast \& Walker 1987; see Sandage, Bell \&
Tripicco 1999).
Cepheids in our sample seem particular easy to spot with high reliability,
however.
We overplot the RGB isochrones of old (12Gy) populations with a range of
reasonable compositions ($Z = 0.0004$, 0.001, 0.004, 0.008 and 0.02) using the
theoretical models of Girardi et al.~(2002).
$R-I$ colors are transformed into $V-I$ using Bessel (1979).
We plot the colors of microlensing candidates found, and their brightest
magnitude consistent with observations in the original (unsubtracted) images,
as explained further below.
Note that according to HIPPARCOS photometry (Perryman et al.~1997), only about
1\% or less stars on the red giant branch blueward of $V-I = 1.5$ show
variability even at the level of a few hundredths of a magnitude, and redward
of this the amplitude rises only gradually, evidently.

\subsection{Microlensing Criteria in $R$ and $I$ and Final Samples}

We now include both $R$ and $I$ in our event discrimination, and produce
several new event criteria, accordingly.
We produce a joint fit in both $R$ and $I$ for the paczynski curve, with seven
fit parameters ($u_0$, $v_0$, $t_0$, baseline flux in $R$, flux offset in $R$,
baseline, baseline flux in $I$, and flux offset in $I$).
The offset flux parameters arise from the possibility that the reference image
includes flux from epochs when the star was not at baseline, hence resulting in
flux subtracted from the star's signal in individual epochs.
The lensing geometry parameters ($u_0$, $v_0$, and $t_0$) are constrained to
the same values in both $R$ and $I$.

\bigskip
Criterion V. Joint $R$ and $I$ $\chi^2/\nu$ threshold:
We set a threshold in reduced $\chi^2/\nu = 1.5$, which will rule out less than
1\% of true lensing events for lightcurves with the number of points as ours,
much less than one event for the sample surviving criteria I - IV.
This criterion actually reduces the surviving candidates to 17 (from 26).
\bigskip

While this evidently reduces the number of variable star contaminants, there
still may be others, such as miras, which are only sampled at a few epochs
while deviating significantly during a secondary maximum peak.
While such an event might be highly discrepant, it might not add enough to the
total $\chi^2$ to push the entire fit beyond the threshold.
With the extra information supplied by the $I$-band data and fit, we should
test again for any repeating or secondary maxima beyond the primary peak.

\bigskip
Criterion VI. No secondary peak in $R$ and/or $I$:
Any contiguous peak (or drop) relative to the baseline in contiguous epochs
(including INT lightcurvepoints) resulting in an increase in $\chi^2$ of more
than 100 will cause the candidate event to be eliminated.
The probability of a true microlensing event showing this behavior is minimal.
This reduces the surviving candidates to 16.

\subsection{Final Selection of Microlensing Events Based on $R-I$ Color}

Note that the ``line of danger'' in Figure \ref{fig:cmd} due to detectable
source variability (outside the cepheid instability strip) corresponds to
$M_I\approx (V-I)_0-6$.
Since $(V-I) \approx (I-H)$ for red giants (see Phillips et al. 1986, and
Neely, Sarajedini \& Martins 2000, for instance), this condition corresponds
to roughly a constant $M_H \approx -6$.
This indicates that a search for near-IR bright sources is likely to be nearly
as effective in eliminating questionable sources as an explicit search for
variability.
We have started such an IR investigation (Kuijken et al.~2002).

We can get an upper limit to the source flux by inspecting the unsubtracted
images for point sources centered on the position of the variable source.
We can estimate the colors of the microlensing candidate sources by taking the 
ratio of excess flux over the baseline in $R$ and $I$.
If candidate events are due to microlensing, there residual flux
should have the same colors as those of the baseline source.
If the event in question is due to microlening, this should indicate the
source baseline color.
Furthermore, since miras and (most) semiregular maxima are many times brighter
than the baseline flux from the same stars, the same is true of these stars:
their residual flux should be very red, similar to that of the unsubtracted
star.
This is particularly the case since the color of miras change little across
the FWHM of the peak, as shown above.
Microlensing of stable sources is likely to arise between $V-I=0.3$, redward of
the aperiodic blue variable locus, to $V-I = 2.8$, where large portions of the
red giant branch still show little variability (see Figure \ref{fig:cmd}).

The ``line of danger'' plus the bounds in $V-I$ for aperiodic blue variables
and red giant variables were then used to isolate the four microlensing events
consistent with non-variable red giant sources (see Figure 7), with all of the
rest of the 16 candidate events landing to the red of this red giant safe zone
by at least $2\sigma$ in $V-I$.
We also attempted to isolate colors of these stars on $HST$ images, as
described next.

While the images used are still crowded, but at 100 times less density of
sources per PSF area than the VATT data, individual stars are seen easily.
In order for these to be useful, the exposures needed to be in two broad bands
centered between 5000\AA\ and 10000\AA\ wavelength, and at least $\sim$100s in
duration.
We searched for these remaining four candidates on such images taken by $HST$,
and unfortunately found none.
\footnote{Based on observations made with the NASA/ESA Hubble Space Telescope,
obtained at the Space Telescope Science Institute, which is operated by the
Association of Universities for Research in Astronomy, Inc., under NASA
contract NAS 5-26555.
These observations are associated with program \#7376, plus archival data
associated with \#4381 and \#8059.}

Properties of the four individual events consistent with the red giant safe
zone are summarized in Table 3, and their
lightcurves and colors are plotted in Figures 6 and 7, respectively.
(Here the first two numerical digits refer to the epoch season in which the
peak is seen e.g., ``97-'' meaning a peak during August 1997 - January 1998,
and the remaining digits refer to an running index we use to count all variable
sources passing Criterion I,
and ``C'' in the last case refers to an event in the nearside, control field.)~
We make a few further comments about their properties:
{\bf 97-1267:} This curve is well-described by a paczynski curve in $R_jt$ and
$I$ with a reduced $\chi^2/\nu = 1.32$ and no excursions from the baseline
approaching more than about $3\sigma$.
This is the reddest of all candidates in residual flux, however, close to our
previously established threshold.
{\bf 97-3230:} This curve is the extreme case in terms our requirement that
the well-sampled lightcurve include the event maximum.
It is fit to $\chi^2/\nu = 1.06$ by a paczynski curve.
{\bf 99-3688:} While only slightly above our $S/N$ threshold for detection of
the event peak, this event is fit with $\chi^2/\nu = 0.88$.
{\bf 99C-1259:} While more poorly sampled, being on the disk far side, this
event is fit with $\chi^2/\nu = 1.32$.
There are a few individual points along the baseline with residuals of
approximately $3\sigma$, but these are always in the vicinity of more points
that are well-fit.
While it is unfortunate that several of these events sit near our previously
selected criterion thresholds, we see no justification for changing these
$post$ $hoc$, and proceed to analyze our sample.

\section {Analysis and Comparison to Microlensing Models}

All four sources for which we have reliable colors and magnitudes land in
regions of the H-R diagram which are likely to be free of detectable
variability, and are consistent with red giant stars, which should represent
the preponderance of source for microlensing events. 
We have no better hypothesis than to elevate these four events to the category
of true microlensing events. 

(We note that for these four surviving sources, we also turn to another
database of the MEGA survey, a series of KPNO 4m/MOSAIC
\footnote{Based on observations made at Kitt Peak National Observatory, a 
division of The National Optical Astronomy Observatories, which is operated by
Associated Universities for Research in Astronomy (AURA), Inc., under a
cooperative agreement with the National Science Foundation.}
exposures taken from 1997
to 2002 encompassing both fields studied here, with emphasis on the years
1999-2002.
This allows us to further extend the baselines beyond that allowed by the VATT,
MDM 1.3m and INT samples, hence further checking if a variable sources might
appear later at the candidate's position, counter to expectations for a
microlensing event.
For three of the four candidates, 97-3230, 99-3688 and 99C-1259, we find
no further variability through 2002, lending support to the possibility of
these being true microlensing events.
For 97-1267, the situation is more complex, since this source is shown to
appear near a fainter, occasionally and slowly varying star.)

Our criteria on sampling (I and III) are sufficiently explicit that we can
compute their effects on our microlensing event detection efficiency given our
knowledge of the sampling time series for our survey, and the other criteria
are sufficiently inclusive of microlensing events as to have negligible impact
on this efficiency. 
We proceed to calculate this efficiency. 
A large number ($1.6\times 10^6$) of lightcurves are generated, based on the
sampling in both the nearside and farside fields. 
The fit parameters are varied over generous ranges. 
Each lightcurve is then tested to see if four consecutive samples exceed
4$\sigma$ above baseline (Criterion III). 
We thus derive a probability of event detection as a function of the fit
parameters, most importantly $t_{1/2}$ and peak flux (expressed as peak
signal to noise). 
This detection efficiency is then folded in to the lensing model of Baltz et
al.~(2003), with source luminosity function and lens mass function applied. 
The time sampling is significantly better for the farside field compared to the
nearside, control field. 
While we see only one probable event on the near side of M31 versus three on
the far side, this is due in part to the difference in sampling.

The model for microlensing rates tells us the distribution of events both as a
function of event FWHM in time $t_{1/2}$ and as a function of position across
the face of M31. 
The latter is striking, in that the distibution of events
across the farside field has a noticibly flatter gradient as one moves out
from the major axis than does the actual distibution of sources.
One might expect this from microlensing events for two
reasons: 1) the lensing cross-section increases roughly linearly with distance
out from the major axis, as explained above, and 2) lensing detection
efficiency is decreased by higher background surface brightness, as seen
closer to the enter of M31. 
This last effect is more than offset by the increase in
density of sources, but the net effect is for event rates to drop as the
square-root of source density rather than linearly (Baltz et al.~2003).
Neither one of these effects should impact the detection of miras, which are
sufficiently bright to be detected across our field, and whose density should
be expected to follow roughly the variation in surface brightness.

We apply the spatial and timescale distribution of these four events to a
series of models varying the fraction of halo dark matter $f_b$ that is
contributed by microlensing objects, and the component mass $m$ of each such
lens. 
For the sake of discussion we assume that $m$ is described by a delta-function
distribution. 
We perform a maximum likelihood analysis comparing the event sample to the
microlensing models. 
This yields the most likely values for $f_b$ and $m$ and the uncertainties
around them. 
For the sake of discussion we pick a halo model from among those computed by
Baltz et al.~that is close to an isotropic, singular, isothermal sphere:
flattening axis ratio $q=1$ and core radius $r_c = 1$~kpc i.e., implying
lensing mass density
\begin{eqnarray}
\rho(x,y,z) = f_b\,\frac{V_c(\infty)^2}{4 \pi G}
\frac{e/(q\,\sin^{-1}e)}{x^2+y^2 + (z/q)^2 +r_c^2}.
\end{eqnarray}

The maximum likelihood calculation performed varying $f_b$ and $m$ shows that
the model where all lensing is due just to the known stellar population of the
bulge and disk (``self lensing''), in which all halo dark matter is non-lensing
on relevant scales, is excluded at about $2.2 \sigma$. 
(We implicitly introduce a positive-definite lensing fraction, since in some
portions of the field even a mildly negative $f_b$ would introduce negative
total rates of microlensing events.)~
This is our primary evidence for the presence of a halo microlensing signal in
M31, and seems to arise primarily due to the shallow gradient across the field
and to some degree the farside/nearside asymmetry. 
If we marginalize over $f_b$ and $m$ in turn, we determine a measurement for
each parameter separately to be microlensing halo mass fraction of
$f_b = 0.29^{+0.30}_{-0.13}$ of the total dark matter, and a lensing component
mass $m$ between 0.02 and 1.5 $M_\odot$ ($1 \sigma$ limits, with a most favored
value of $0.53~ M_\odot$). 
This component mass constraint is very weak, especially considering that we
adopted a prior of $10^{-3}~M_\odot < m < 10~M_\odot$ in constructing our
maximum likelihood estimate.
This fit to $f_b$ adequately predicts the number of events (see, for
instance, Figure 8), and accounts successfully for the events that do survive
being found at large radii, as we further treat below.

Another term not included explicitly in our detection efficiency analysis, is
the loss of point source detections as potential microlensing events due to the
crowding of residual point sources in the subtracted frames.
This can lead to the rejection of a true microlensing event due to the apparent
variation in the baseline (Criterion VI) or overall bad fits in $\chi^2$
(Criteria II, IV and V).
This effect is legislated by the diameter of the photometric aperture used for
determining flux from a given residual point source, three times the FWHM of
the final PSF.
We investigated this effect briefly by inspecting the density of residual point
sources on the subtracted images and also, by evaluating the loss of synthetic
residual sources introduced into the image, we attempted to define a boundary
in the image where the loss due to adjacent variable sources reached a set
fractional value.
The value for halo fraction $f_b$ quoted above ($0.29^{+0.30}_{-0.13}$) is set
by including all points in the sample beyond the boundary at which the number
of sources rejected due to confusion with adjacent variable sources no longer
dominated by those retained.
In other words, we estimate the value of $f_{lost} = N_{lost}/N_{tot}$, where
$N_{lost}$ is the average number of residual point sources neglected, and
$N_{tot}$ the total number introduced.
For $f_b = 0.29^{+0.30}_{-0.13}$, this corresponds to $f_{lost} < 0.3$
everywhere in the sample, including at the worse case along the inner boundary,
corresponding roughly to line running parallel to the major axis at a distance
4 arcmin away, and a circle of radius 5 arcmin centered on the nucleus
(whichever of these two curves is farther out).
Expanding this region to where $f_{lost} < 0.01$ increases $f_b$ by only 0.01
(with similar error bars).
The overall sensitivity of our remaining survey sample as a function of mass is
shown in Figure \ref{fig:sensitivity}.

Our constraint on the microlensing halo mass fraction is subject to several
caveats, most seriously its dependence on a given halo model.
The (nearly) singular isothermal sphere model allows the most direct comparison
to Galactic results (Alcock et al.~2000a) but does not account for the
different paths through the halos probed by the two surveys.
Our survey path passes largely to the interior to that towards the LMC, so
will suffer a relative loss in halo microlensing signal given $r_c > 0$.
Even for a moderate $r_c = 5$~kpc, the microlensing optical depth is reduced
versus a singular model by a factor $\sim$1.5 (Gyuk \& Crotts 2000), increasing
our result potentially even more compared to the MACHO Galactic result.
Furthermore, we select only for simple point lenses in requiring good
paczynski (or gould) fits.
This result excludes contributions from binary and planetary lens masses.
Several binary lenses have been found (Alcock et al.~2000b,
Udalski et al.~1994), as have binary source
events (Alcock et al.~2001); either type (especially the former)
might produce events eliminated by our current filters.
Estimating the number of binary events in the Galactic sample might be
complex (see Di Stefano \& Perna 1997), but judging by the
radically non-paczynski binary lightcurves in the Galactic sample, we
might easily expect to have overlooked 10\% of the lensing optical
depth in M31's halo.

\section{Discussion}

How do our results compare to those from Galactic Halo microlensing searches?
The most detailed results come from the MACHO survey, reporting (Alcock et
al.~2000a) that, during 5.7 years of observations, the detection of 13-17
events towards the LMC when only $\sim$2-4 events would have been expected due
to known, intervening stellar populations. 
This corresponds to a microlensing optical depth, $\tau_{LMC} =
1.2_{-0.3}^{+0.4} \times 10^{-7}$ for $2 < \hat{t} < 400$ days. 
The events have timescales of $<\hat{t}>$ = 34-230 days and the most probable
mass for the events is $<m>$ = 0.15-0.9 \msolar. 
The maximum likelihood halo fraction is $f_{b} $ = 20\% (8\% - 50\%,
95\% confidence interval) implying a total mass in MACHOs of M$_{\rm MACHO}$ =
$9_{-3}^{+4} \times 10^{10}$ \msolar~ ($r <$ 50 kpc). 
The EROS survey reports a result consistent with the central MACHO value, but
also consistent with no Galactic microlensing halo (Afonso et al.~2003).
In addition to searching for events with well-sampled, long-duration
microlensing lightcurves, the MACHO \& EROS groups conducted an
analysis (``spike analysis'') in which they searched for very
short timescale brightenings in order to place limits on low mass MACHOs. 
Their conclusion is that Milky Way halo dark matter cannot be comprised of
objects in the mass range $2.4 \times 10^{-7}$ \msolar $< m <$ $5.2 \times
10^{-4} M_\odot$ (Alcock et al.~1996).

The M31 halo microlensing fraction $f_b$ we find is consistent with that seen
by MACHO towards the LMC.
Our central value is higher (by $0.7\sigma$ using only our error, or $\approx
0.4\sigma$ combining both surveys' errors).
We cannot argue for any inconsistency between the two surveys' results.

The positive, marginally significant halo signal we report is due in large
part to the asymmetric distribution of events across the face of M31,
slightly favoring the far side as would be expected from a microlensing dark
matter halo.
Our result is approximately as consistent with no halo as that reported by
MACHO, however.
On the basis of further M31 microlensing evidence, however, we tend to accept
the positive halo indication.
The VATT/Columbia survey serves as the pilot study for a larger survey, MEGA
(``Microlensing Exploration of the Galaxy and Andromeda:'' Crotts et al.~2001),
and the first results from this effort (de Jong et al.~2003) also shows a
marginally significant farside surplus asymmetry.
While this other work does not estimate $f_b$, such an asymmetry is a nearly
unique marker of a dark matter halo of microlensing objects.
Later data from this survey will encompass almost an order of magnitude more
observations, so indicates that the result will become more clear in the near
future.

Perhaps the most surprising conclusion drawn from the MACHO data is that the
lenses lie in a mass range occupied by stellar objects and are well above the
hydrogen burning limit ($\sim$0.065 \msolar: Chabrier \& Baraffe [2000]) which
defines the brown dwarf/stellar boundary. 
Creating plausible models which can explain these results without violating
other astrophysical constraints has proven to be quite a challenge for
theorists. 
The only candidates in this mass range to be excluded by direct observation are
low-mass stars (Boeshaar, Tyson \& Bernstein 1994; Graff \& Freese 1996) and
the most viable baryonic candidates
are a population of old white dwarfs (WD) which have been previously overlooked
because their colors are bluer than expected due to an increase in atmospheric
H$_{2}$ opacity (Hansen 1999).
Various surveys (Ibata et al.~1999, 2000; Oppenheimer et al.~2001a, b) have
now discovered a significant number of such high-proper motion, faint WDs whose
kinematics appear to be consistent with those of Galactic halo objects. 
Alternatively, there are those who would argue that these events are not
associated with a dark matter halo but are, rather, explained as lensing by an
intervening population of stars along the line-of-sight to the Magellanic
Clouds (see Zaritsky \& Lin 1997, Zhao 1998) or as self-lensing by stars
within the LMC/SMC (Sahu 1994).
As an alternative, we point out that primordial black holes
with masses of $\sim$1 \msolar\, can be formed from density perturbations
created during the quark-hadron phase transition in the early universe
(Crawford \& Schramm 1982, Jedamzik 1997). 
However, although initially baryonic these black holes are classified as
non-baryonic dark matter candidates because they form before the epoch of BBN
and, therefore, are not subject to its constraints.

Comparing the current survey to MACHO, their constraints on $m$ is somewhat
more restrictive.
They argue for masses above the 0.07$M_\odot$ hydrogen core-burning threshold,
at a level of certainty of about 90\%.
While in M31 we cannot rule out the possibility that this represents a
population of red main-sequence dwarf stars, in our own Galaxy this is ruled
out.
While our current survey adds little new information as to the nature of these
objects, it tends to confirm that the population is seen to exist now in
two galaxies, and is likely to be a universal phenomenon.
With galaxy halo dark matter accounting for nearly the same fraction of
universal closure density as baryons (Rubin 1993), with the WMAP value now
set at $\Omega_B = 0.046$ (Spergel et al.~2002).
We are speaking of a contribution of about 1\%, well within the baryon
budget.

An M31 microlensing result might be contaminated by foreground lenses, as has
been suggested for the LMC.
Since the timescales of Galactic and M31 halo events are similar, given the
same lens mass, an M31 survey is susceptible to foreground Galactic halo
lensing.
Extrapolating previous results (Zhao 1998), an optical depth of $10^{-7}$ with
a sheet of matter at 10 kpc from the source (or observer), implies a surface
density $\sim 15 M_\odot/pc^2$.
Our signal implying a microlensing halo corresponds to an optical depth roughly
an order of magnitude larger than this.
In the Milky way, just judiciously covering the MDM farside field with
such a sheet would only require $3\times 10^5M_\odot$.
In M31, this requires $\sim 10^9 M_\odot$.
The latter would imply an unusually massive tidal stream, whereas such a
possibility in our Galaxy is perhaps reasonable if one ignores the
double coincidence of its appearance both in front of the MACHO and current M31
survey fields.
For either galaxy, a spherical shell would need about $\sim 
10^{11}M_\odot$ at 10 kpc to make the optical depth, with mass at other
locations scaling as the radius: $M_{shell} \approx 10^{11} M_\odot
( D / 10~kpc )$.
This approaches a simple re-creation of the original halo lensing mass problem.

From this survey we have insufficient data to state whether these lenses
arise in a thick disk or a true halo.
To do so will require more events, scattered over a larger portion of the face
of M31.
Fortunately, such a survey is practible (Baltz et al.~2003) and is currently
underway (Crotts et al.~2001).
While the results presented here are interesting in their implications for the
universality of disk galaxy halo dark matter, we would prefer to have better
sampled lightcurves, to cover more of the face of M31 in order to get better
leverage on the spatial variation of the lensing population, and to better
catalog populations of variable stars in the same field which might
masquerade as microlensing events.
We look forward to additional progress in using M31 to generalize and extend
the insight which has been won from microelensing searches in nearby galaxies.

\acknowledgements

We wish to thank Andrew Gould and collaborators for providing data from their
observations as part of a cooperative agreement with our group to observe M31
at the MDM 1.3-meter telescope.
E.B. acknowledges support from the Columbia University Academic Quality Fund.
A.C. was supported by grants from NSF (AST 00-70882, 98-02984 and 95-29273 and
INT 96-03405) and \mbox{STScI (GO-7376)}.
We are grateful to Patrick Cseresnjes and David Alves for helpful discussions.

\pagebreak
 
\begin{table}
\caption[Survey Field Positions]{Celestial coordinates (J2000.0) for the centers of our survey fields.  Fields are rotated by $37^{\circ}.7$.}\label{tbl:skycoo}
\begin{center}
\begin{tabular}{l c c}
\hline
\hline
Telescope+Field & $\alpha_{2000.0}$ & $\delta_{2000.0}$ \\
\hline
VATT Target &00$^{h}$ 43$^{m}$ 16$^{s}.5$ &+41$^{\circ}$ 11\arcmin ~33\arcsec\\  
VATT Control&00$^{h}$ 42$^{m}$ 13$^{s}.4$ &+41$^{\circ}$ 20\arcmin ~44\arcsec\\ 
\hline
MDM Target  &00$^{h}$ 43$^{m}$ 28$^{s}.1$ &+41$^{\circ}$ 11\arcmin ~48\arcsec\\
MDM Control &00$^{h}$ 42$^{m}$ 17$^{s}.8$ &+41$^{\circ}$ 22\arcmin ~44\arcsec\\
\hline
\end{tabular}
\end{center}
\end{table}

{\tiny
\begin{table}
\caption{Properties of the MDM \& VATT nightly image stacks}
\begin{center}
\begin{tabular}{c c c c|c c c c}\hline\hline
UT Date&Obs&Times[s]$^*$&FWHM[\arcsec]$^{**}$ &
UT Date&Obs&Times[s]$^*$&FWHM[\arcsec]$^{**}$ \\
\hline
1996-09-20&VATT&7200/-/-/-         &1.26&1997-12-04&MDM& 3600/1200/-/-   &1.76\\
1996-09-21&VATT&4800/6000/-/-      &1.04&1997-12-05&MDM& 5400/1200/-/-   &1.67\\
1996-09-22&VATT&3600/3600/4800/-   &1.04&1998-09-29&MDM& 9000/-/7200/1800&1.27\\
1996-09-23&VATT&6600/-/-/-         &0.93&1998-10-02&MDM&11160/9000/8400/-&1.51\\
1996-10-13&VATT&11400/6600/-/-     &1.05&1998-10-03&MDM&13680/7800/-/9600&1.68\\
1996-10-15&VATT&13120/4600/-/-     &1.55&1998-10-05&MDM&10802/9000/3300/-&1.76\\
1996-10-16&VATT&6600/2400/2400/600 &1.14&1998-10-06&MDM& 8280/4200/-/1500&1.99\\
1996-10-17&VATT&10800/3600/2400/-  &1.16&1998-10-08&MDM&10800/4200/-/-   &1.59\\
1996-10-18&VATT&1200/-/-/-        &1.20&1998-10-09&MDM&11520/6000/900/600&1.53\\
1996-10-20&VATT&2400/400/-/-       &1.90&1998-10-11&MDM&11520/4200/-/-   &1.53\\
1996-10-21&VATT&5400/2400/2700/-  &1.30&1998-10-28&MDM&10800/3600/-/12000&1.50\\
1996-10-27&VATT&2400/-/-/-         &2.55&1998-10-29&MDM&10800/4200/9000/-&1.43\\
1996-11-11&VATT&1800/-/-/-        &1.36&1998-10-31&MDM& 3600/3000/-/10800&1.96\\
1996-11-12&VATT&9900/3600/-/-      &0.98
&1998-11-01&MDM&10800/9000/10800/-$^{***}$&1.41\\
1996-11-13&VATT&10800/3600/-/-     &1.09&1998-11-03&MDM&10800/9000/-/-   &1.46\\
1996-11-16&VATT&9900/2400/2100/-&1.91&1998-11-04&MDM& 7920/7200/1200/1200&1.52\\
1997-09-12&MDM& 3000/-/-/-         &1.45&1998-11-06&MDM& 9360/9000/-/-   &1.38\\
1997-09-13&MDM&14400/-/-/-        &1.55&1998-11-07&MDM&10080/5400/900/600&1.84\\
1997-09-14&MDM& 2400/-/-/-         &3.16&1998-11-09&MDM& 7920/4200/-/300& 1.65\\
1997-09-15&MDM& 3600/-/-/-         &1.63&1998-11-10&MDM&10080/4200/1200/-&1.76\\
1997-09-23&MDM& 3600/-/-/-         &1.53&1998-11-13&MDM& 3600/1200/1800/-&2.03\\
1997-09-27&MDM& 2520/1440/-/-      &1.47&1998-11-15&MDM& 7200/3600/-/-   &1.61\\
1997-09-28&MDM& 5760/4200/-/-      &1.95&1998-11-16&MDM& 8280/3600/7200/-&1.46\\
1997-09-29&MDM& 2520/1800/-/-      &1.41&1998-11-18&MDM& 8640/3300/1200/-&1.53\\
1997-09-30&MDM& 6480/-/-/-         &1.46&1999-08-30&MDM& 2160/-/-/-      &2.20\\
1997-10-01&MDM& 7200/1440/-/-      &1.44&1999-09-15&MDM&10080/3600/-/-   &1.50\\
1997-10-04&MDM& 5760/3780/3000/1200&1.89&1999-09-16&MDM&13320/3600/-/-   &1.35\\
1997-10-05&MDM& 4680/3000/2400/1380&1.87&1999-10-15&MDM& 6840/2400/-/-   &1.79\\
1997-10-06&MDM& 5400/3600/-/-      &1.87&1999-10-16&MDM& 2160/1200/-/-   &1.50\\
1997-10-07&MDM& 9360/4200/-/-      &1.78&1999-10-17&MDM& 7560/4200/-/-   &1.98\\
1997-10-08&MDM& 2160/-/-/-         &2.20&1999-10-19&MDM& 9720/5100/-/-   &1.54\\
1997-10-17&MDM& 5400/4200/-/-      &1.65&1999-10-20&MDM&10800/5400/-/-   &2.23\\
1997-10-19&MDM& 3960/3000/-/-      &1.55&1999-11-30&MDM& 7560/3600/-/900 &1.93\\
1997-10-20&MDM& 5040/6600/-/-      &1.29&1999-12-01&MDM& 2160/-/1200/60  &1.72\\
1997-11-08&MDM& 5760/3000/-/-      &1.45&1999-12-03&MDM& 6480/2400/-/600 &1.90\\
1997-11-09&MDM& 7200/4800/-/-      &1.56&1999-12-08&MDM& 5040/3000/-/-   &1.86\\
1997-11-10&MDM&10080/6600/-/-   &1.51&1999-12-09&MDM& 4320/2700/7200/1800&2.40\\
1997-11-15&MDM&9000/7200/10800/-&1.87&1999-12-12&MDM&8280/3000/10800/1800&1.45\\
1997-11-16&MDM& 8640/2400/-/-      &1.51&1999-12-14&MDM& 6840/2400/-/-&   3.09\\
1997-11-17&MDM& 5400/4800/-/-   &1.74&1999-12-15&MDM& 8640/4800/2400/4800&1.70\\
1997-11-18&MDM& 8640/2700/4800/5400&1.94&1999-12-19&MDM& 3240/-/-/-      &2.06\\
1997-11-19&MDM& 8280/2400/-/-    &1.54&1999-12-20&MDM& 6480/4200/2400/600&1.90\\
1997-11-29&MDM& 9000/1800/-/-      &1.58&1999-12-28&MDM& 5760/1800/-/-   &1.74\\
1997-11-30&MDM&10440/3000/-/-      &1.76& & & & \\
\hline
\multicolumn{8}{l}{$^*$ Total exposure time for $R_{JT}$ (Target), $I_c$
(Target), $R_{JT}$(Control), $I_c$ (Control), respectively.} \\
\multicolumn{8}{l}{$^{**}$ Image FWHM for $R_{JT}$; $I_c$ is usually better by
$\sim$10\%.} \\
\multicolumn{8}{l}{$^{***}$ In order to shorten the length of this table we
have combined some entries from adjacent nights onto the same} \\
\multicolumn{8}{l}{ line when they
involve different filter/field combinations.} \\
\end{tabular} 
\end{center}
\end{table}
}

\begin{table}
\caption[]{Properties of Candidate Microlensing Events}
\begin{center}
\begin{tabular}{l c c c c c}\hline\hline
Events ID& RA~ (J2000)~ Dec       & MJD of Peak & $R_{diff}$ & $t_{fwhm}$ [d]\cr
\hline
97-1267  &00 43 22.87~~ $+$41 05 30.0& 50765.2  & 22.2       & 26.5 \cr
97-3230  &00 43 02.90~~ $+$41 07 14.5& 50715.8  & 20.3       & 17.3 \cr
99-3688  &00 43 57.27~~ $+$41 11 56.3& 51468.3  & 21.8       & 2.2  \cr
99C-1259 &00 41 54.16~~ $+$41 21 40.9& 51519.5  & 20.3       & 10.3 \cr
\hline
\end{tabular}
\end{center}
\end{table}

\begin{figure}
\epsfig{file=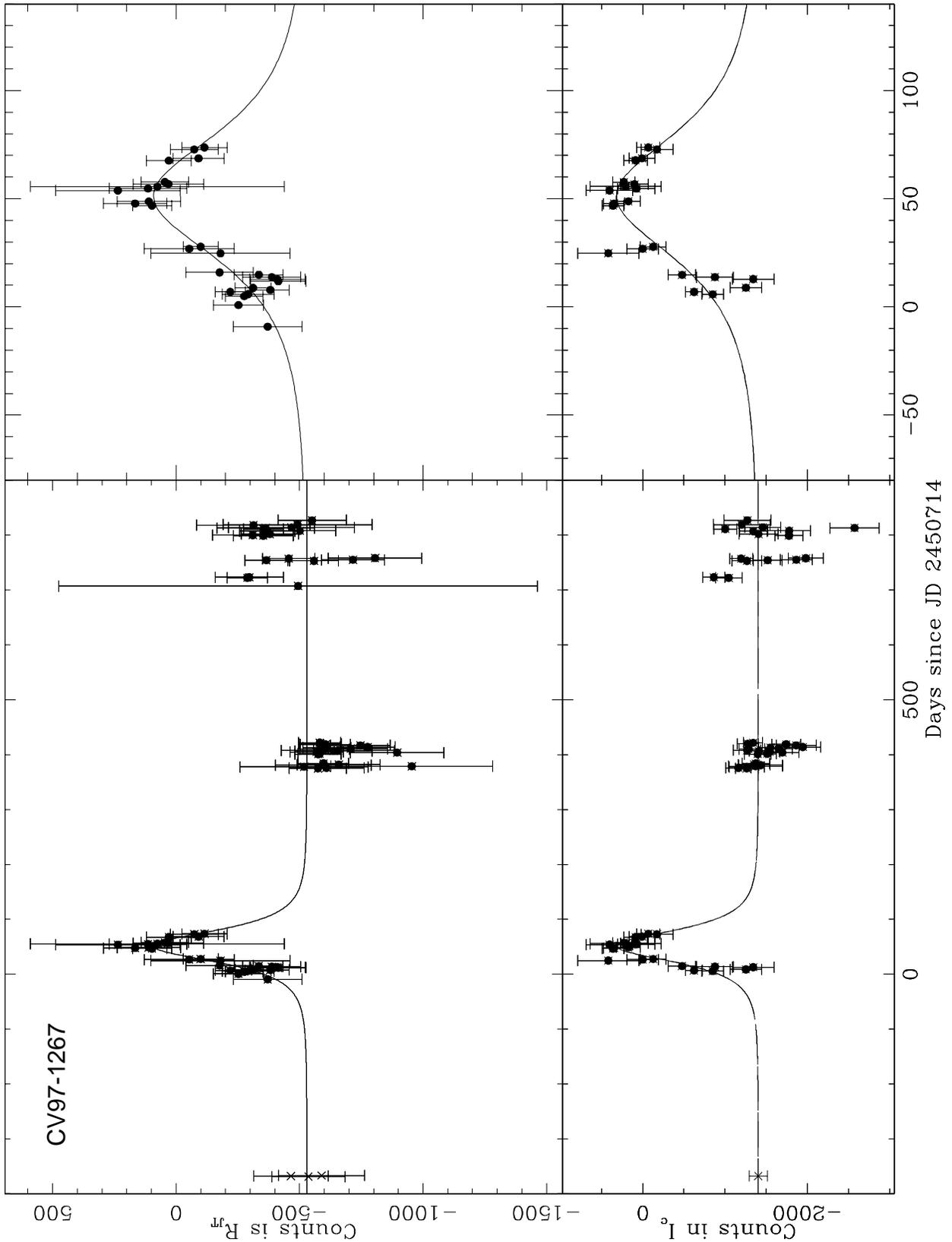,width=\columnwidth}
\vskip -0.6in
\caption{
(This and next 3 pages):
Candidate lightcurves from the total sample, plotted in ADU counts
in the $R$ and $I$ bands as a function of Julian date.
Points used in the paczynski fits (from the MDM 1.3-meter) are indicated by
the solid symbols; data from the VATT (first season) and INT (last season) are
indicated by small diagonal crosses.
}
\label{fig:full}
\end{figure}

\begin{figure}
\epsfig{file=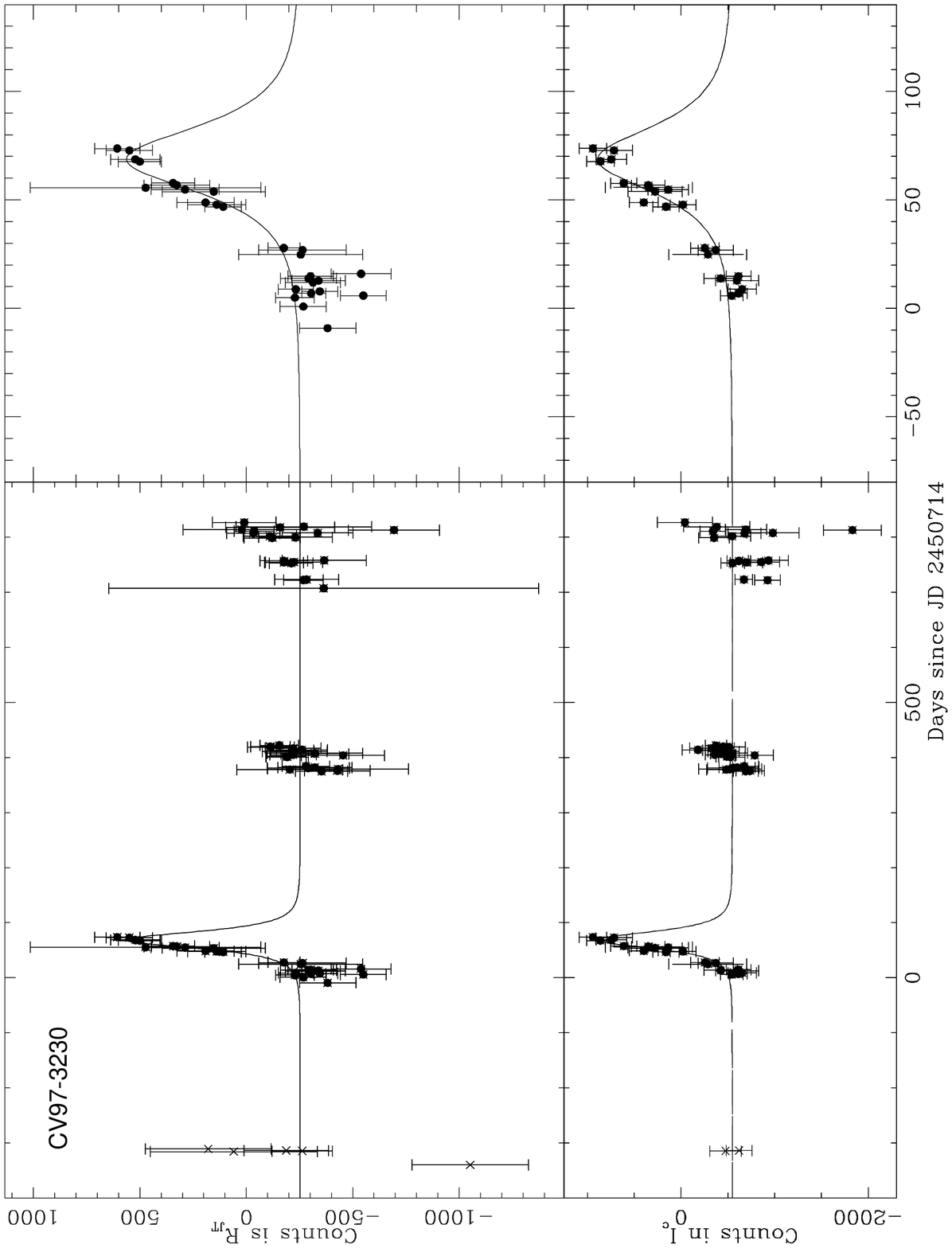,width=\columnwidth}
\end{figure}

\begin{figure}
\epsfig{file=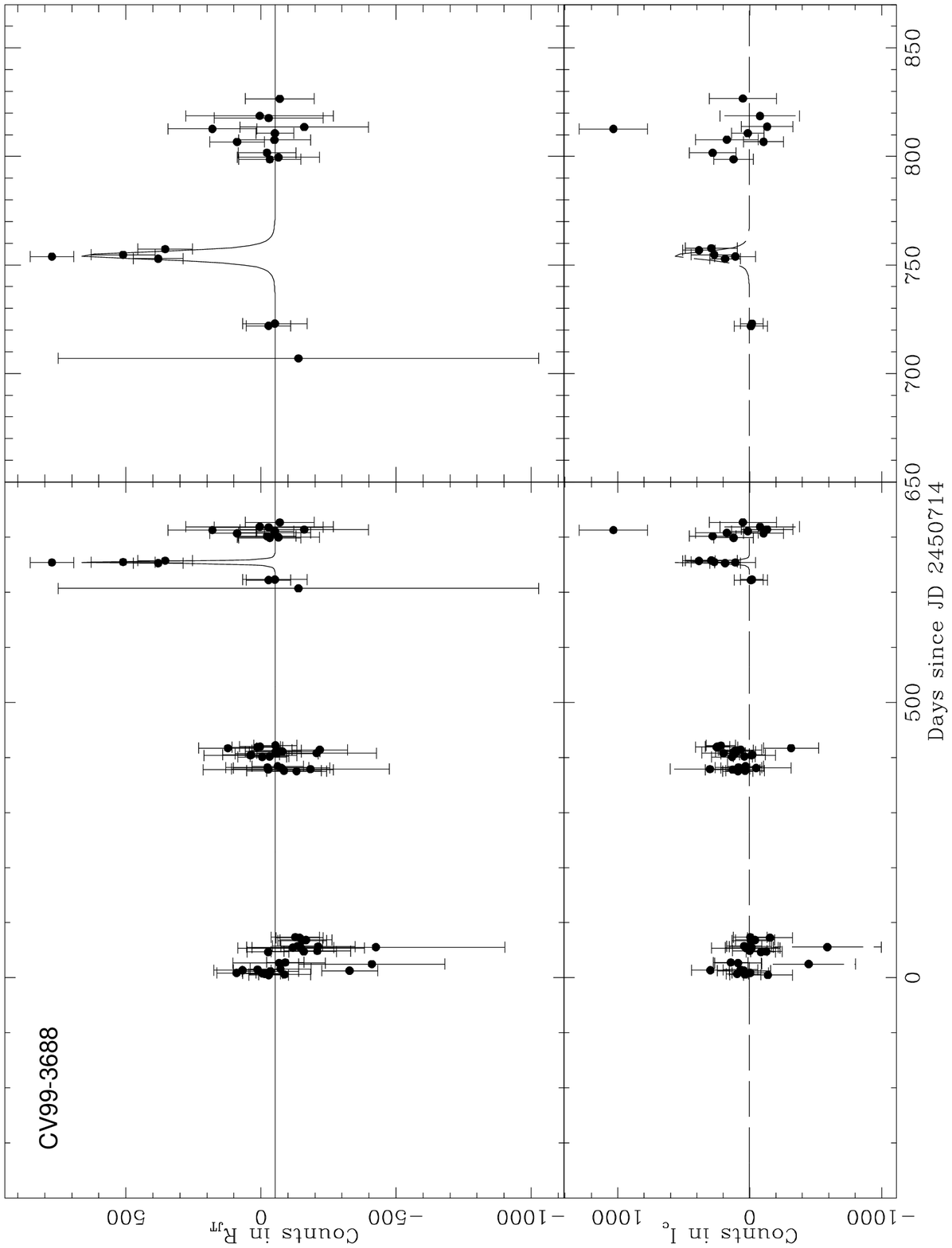,width=\columnwidth}
\end{figure}

\begin{figure}
\epsfig{file=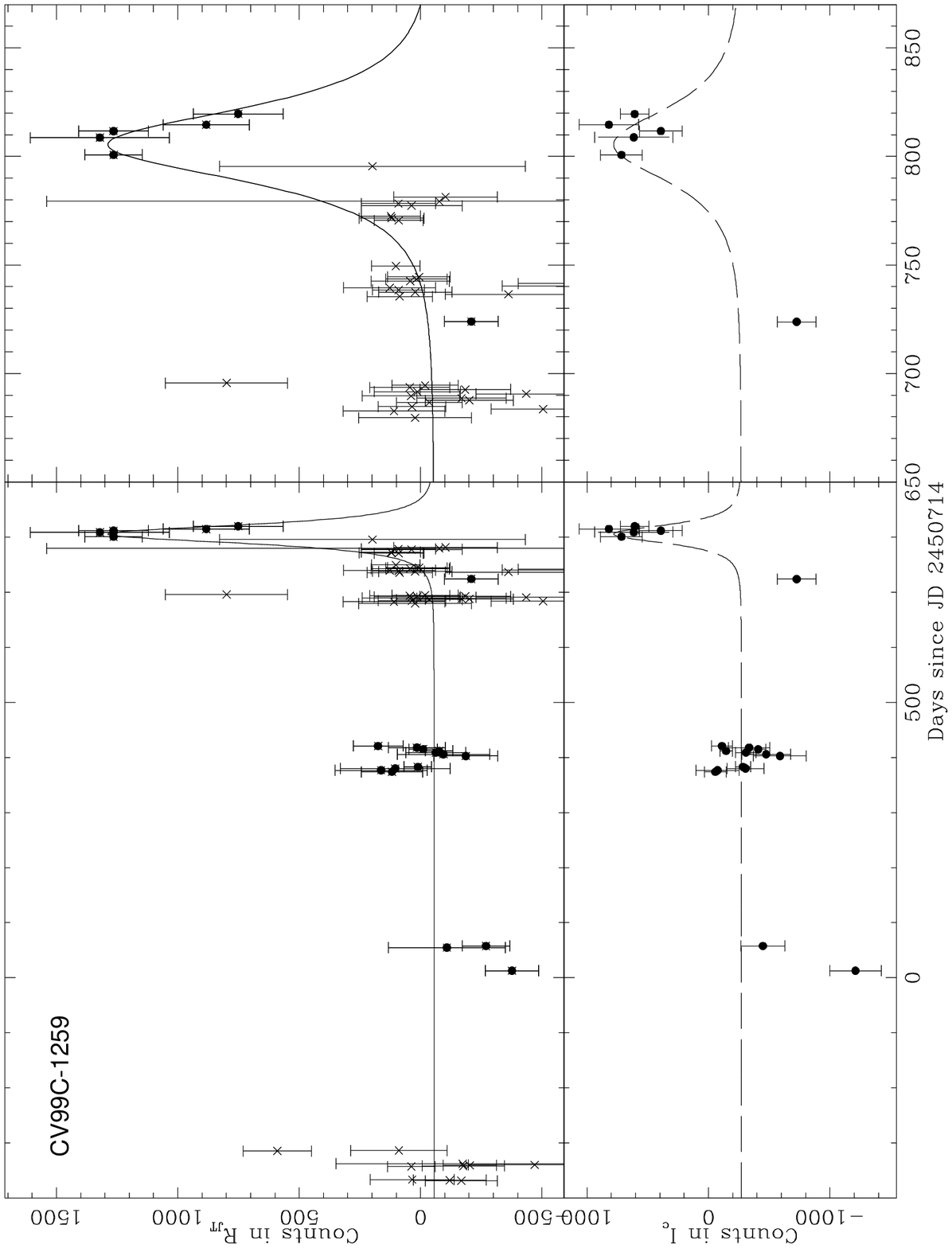,width=\columnwidth}
\end{figure}

\begin{figure}
\epsfig{file=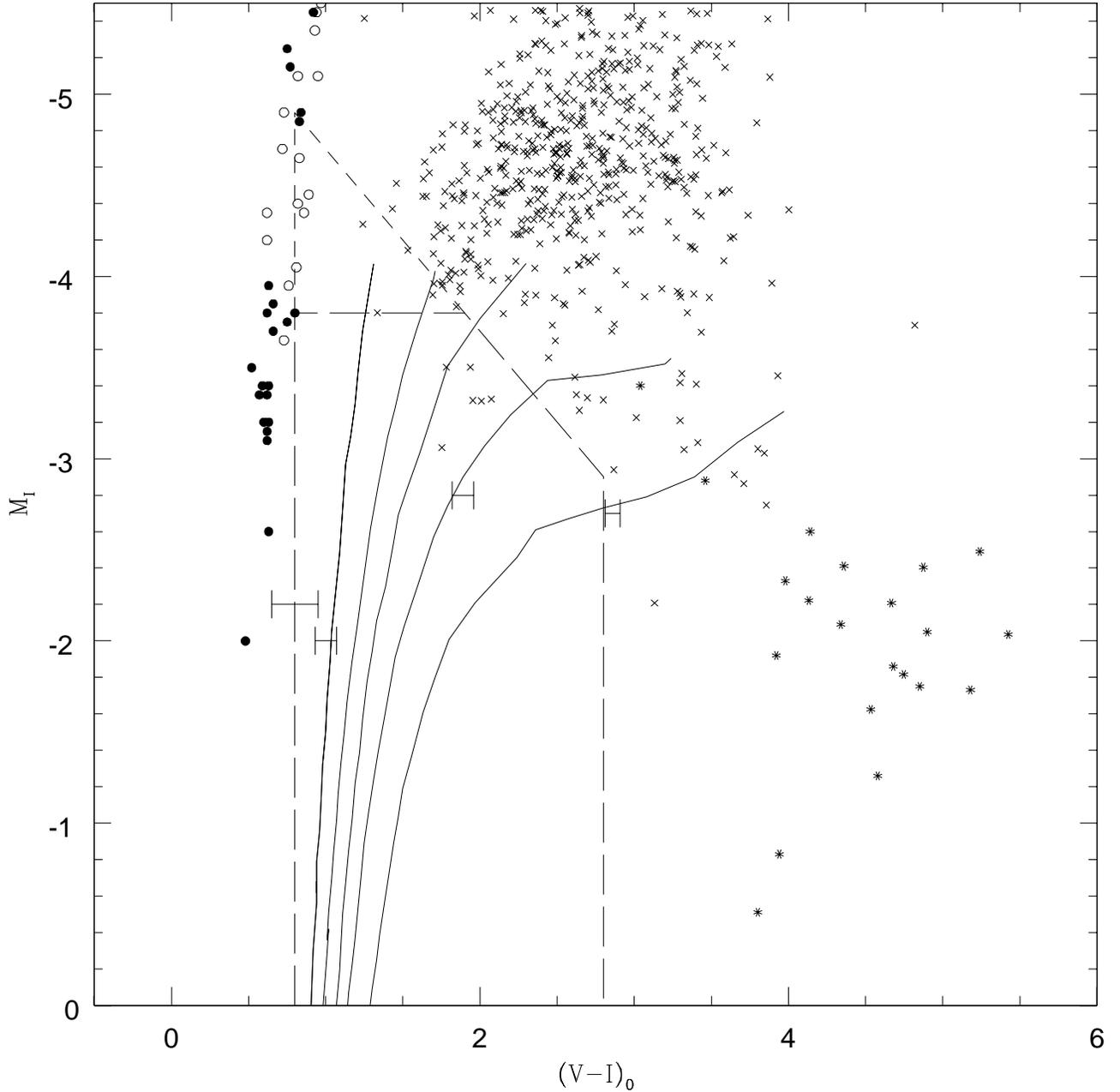,width=\columnwidth}
\caption{Color-magnitude diagram in $V$ versus $I$ is shown for several
collections of stars.
The errorbars indicate the $\pm 1 \sigma$ color limits of the four surviving
microlensing events (from right to left: 99-3688, 99C-1259, 97-3230 and
97-1267) that have passed all seven criteria.
The vertical positions indicate the upper limit (roughly 95\% confidence) on
$M_I$ from inspection of the unsubtracted images taken along the baseline.
Five curves correspond to the red giant branch locii predicted for a range of
internal compositions as might be found in our fields, as calculated by
Girardi et al.~(2002).
The large number of small star symbols indicate (non-cepheid) variables from
the MACHO database that might be expected to rise above our event detection
threshold, as indicated by an event fluence taken from the product of their
fluctuation amplitude, baseline flux, and variability timescale.
LMC variables are indicated by 4-pointed stars and Bulge variables by 8-pointed
stars.
The small circles indicate cepheids: open from the LMC, and filled from the
Galaxy.
A ``safe zone'' of low source variability is defined by cuts in $V-R$ and
a restriction $M_I > (V-I)-5.9$.
(We arbitrarily also remove the region about $M_I = -3.7$.)~
All four surviving microlensing events land on or close to the red giant
branch, consistent with this safe zone region of undetectable intrinsic
variability.
}
\label{fig:cmd}
\end{figure}

\begin{figure}
\epsfig{file=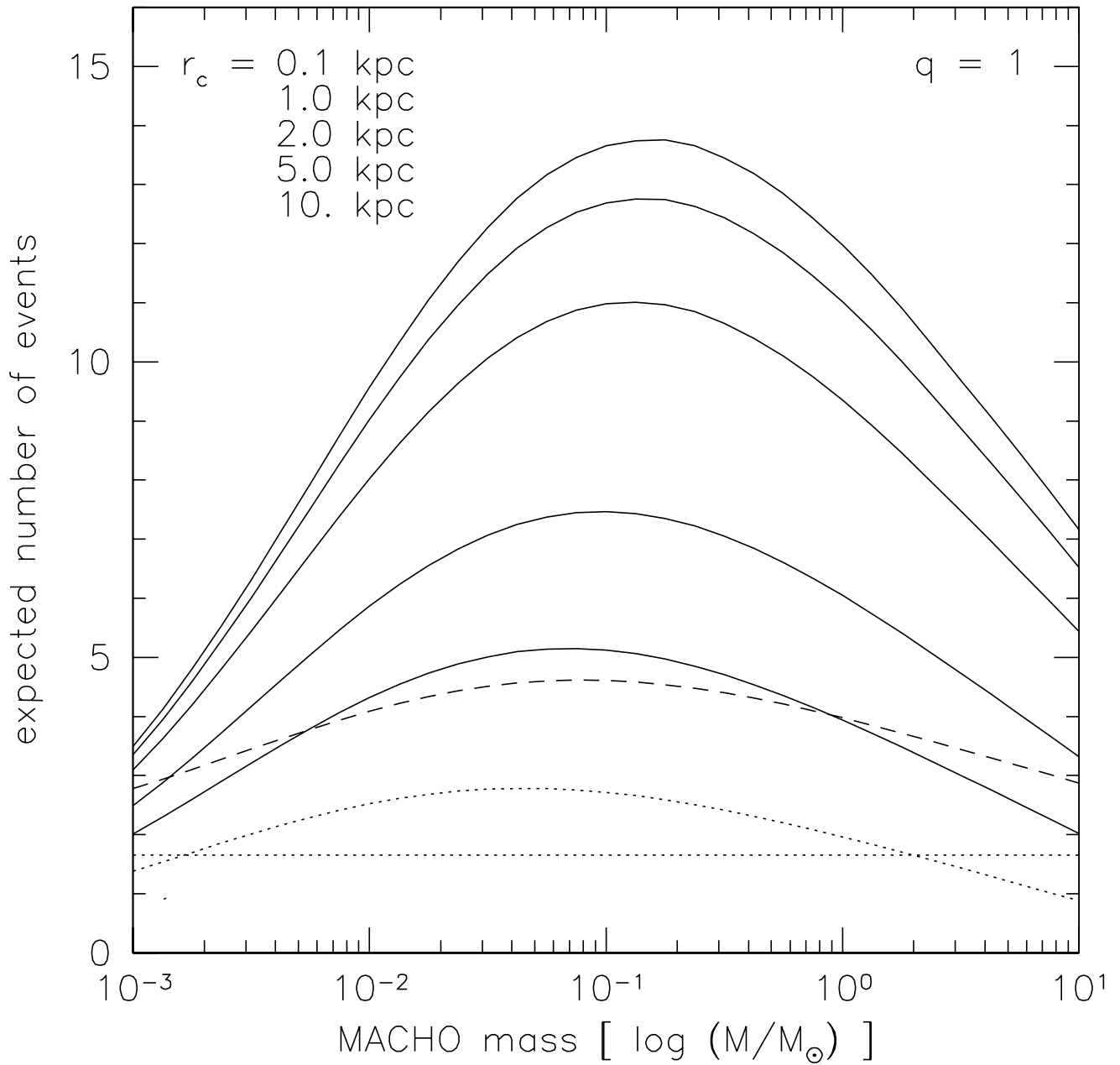,width=\columnwidth}
\caption{
The sensitivity of our survey (for the intermediate case in efficiency effects
of point source confusion) for several types of microlensing populations.
The decreasing solid curves correspond to a 100\% MACHO halo, for increasing
core radius $r_c$, as a function of $\delta$-function lens mass $m$.
The straight dotted line shows the ``self-lensing'' stellar contribution as
defined by a Chabrier mass function (which does not depend on halo component
mass $m$).
The dotted curve shows the contribution from 100\% of Galactic halo dark matter
composed of lenses of mass $m$.
The dashed curve shows the total of self-lensing, and a 20\% Galactic and M31 
contribution.
}
\label{fig:sensitivity}
\end{figure}

\end{document}